\documentclass[iop]{emulateapj}

\usepackage{graphicx}
\usepackage{url}

\shorttitle{Multi-band polarization of Type IIP supernovae due to light echo from circumstellar dust}
\shortauthors{Nagao et al.}

\begin{document}

\title{Multi-band polarization of Type IIP supernovae due to light echo from circumstellar dust}

\author{Takashi Nagao\altaffilmark{1,4,5}, Keiichi Maeda\altaffilmark{1}, Masaomi Tanaka\altaffilmark{2,3}
}

\altaffiltext{1}{Department of astronomy, Kyoto University, Kitashirakawa-Oiwake-cho,
  Sakyo-ku, Kyoto 606-8502, Japan}
\altaffiltext{2}{Astronomical Institute, Tohoku University, Aoba, Sendai 980-8578, Japan}
\altaffiltext{3}{National Astronomical Observatory of Japan, Mitaka, Tokyo 181-8588, Japan}
\altaffiltext{4}{Email: nagao@kusastro.kyoto-u.ac.jp}
\altaffiltext{5}{Research Fellow of Japan Society for the Promotion of Science (DC2)}

\begin{abstract}
Type IIP supernovae (SNe IIP) often show relatively high continuum polarization ($\sim 1$\%) in the late phase. This polarization feature is generally believed to be due to an inner aspherical core revealed in the late phase, while this polarization feature can also be contributed by the effect of polarized-scattered echoes by circumstellar (CS) dust around the SN. In this paper, we propose a unique method to distinguish polarization from the SN ejecta and from the light echo. We quantitatively examine wavelength dependence of the polarization created by the scattered echoes for various geometries and amounts of CS dust. It is found that the polarization in the $U$-band has characteristic features, i.e., the polarization emerges at an earlier phase with higher polarization degree than that in longer wavelengths. These are due to the rapid evolution of the $U$-band light curve as well as higher optical depth of dust in shorter wavelengths. Except for the $U$ band, the polarization increases after the plateau phase, and the polarization degree is generally higher for shorter wavelengths. These polarimetric features can be easily distinguished from the polarization expected from an aspherical core, which predicts almost no wavelength dependence. Moreover, we show that multi-band polarimetric observations for SNe IIP can constrain a parameter space in the CS dust mass and distance from the SNe. We thus encourage multi-band polarimetric observations for SNe IIP.
\end{abstract} 


\section{Introduction}

Type IIP supernovae (SNe IIP) belong to the most common class of core-collapse SNe, which are believed to originate from massive stars ($\gtrsim 8 M_{\odot}$). They play important roles in the universe, both chemically polluting interstellar space and inducing star formation. However, the explosion mechanism is still an open question. It has been widely accepted that one-dimensional simulations cannot reproduce the SN explosions \citep[e.g.,][]{Rampp2000, Liebendorfer2001, Thompson2003, Sumiyoshi2005}. Recently, there are an increasing number of numerical examples leading to an explosion, at least for some progenitor models, found only in multi-dimensional simulations \citep[e.g.,][]{Buras2006, Marek2009, Suwa2010, Muller2012, Takiwaki2012, Bruenn2013, Hanke2013, Couch2014, Takiwaki2014, Lentz2015, Melson2015, Muller2016, Roberts2016}. While these simulations do not yet reach to the explosion energy comparable to the observationally inferred typical value ($\sim 10^{51}$ erg), the multi-dimensional effect is generally believed to be essential in the SN explosion mechanism \citep[e.g.,][]{Maeda2008, Tanaka2017}.

SNe IIP show a unique feature in their continuum polarization: Their continuum polarization rapidly increases ($\sim 1$ \%) just after entering the nebular phase, following generally a small polarization level ($\sim 0.1$ \%) in the plateau phase \citep[e.g.,][]{Jeffery1991, Leonard2001a, Leonard2001b, Leonard2006, Chornock2010, Kumar2016}. This polarimetric behavior could be explained as a highly asymmetric core revealed in the nebular phase (hereafter, the aspherical core model; e.g, Dessart \& Hillier 2011). However, scattering of SN light by aspherically distributed circumstellar (CS) dust is another possibility to produce the net continuum polarization in SNe IIP \citep[hereafter, the dust scattering model; e.g.,][; hereafter, Paper I]{Wang1996, Nagao2017}. In the dust scattering model, the net polarization is produced by aspherically distributed CS dust, which is important information to understand mass-loss processes in massive stars, especially those just before their explosions. Thus, it is important to understand the origin of the polarization of SNe IIP, for understanding not only the explosion mechanism but also the mass-loss processes.
 
One of the biggest differences between the predictions from the aspherical core and dust scattering models is the wavelength dependence of polarization expected in each model. In the aspherical core model, no significant wavelength dependence of the polarization is expected in the nebular phase. This is because the optical depth and the albedo in the H-rich scattering atmosphere is determined mainly by electron scattering, whose opacity is gray, although bound-bound, bound-free and free-free absorption may contribute to the opacity and create some wavelength dependence for the polarization \citep[see, e.g.,][for the wavelength dependence though in the earlier (plateau) phase]{Dessart2011}. On the other hand, the wavelength dependence is generally expected in the dust scattering model (see Section 4.4 in Paper I for details). In Paper I, we examine the effects of the scattered echoes on the SN polarization, for a certain optical wavelength. In this paper, we examine the wavelength dependence of the polarization in the dust scattering model. This paper is structured as follows: In Section 2, we describe our methods. In Section 3, we summarize our results. The paper concludes in Section 4.

\section{Methods}
We perform three-dimensional radiative transfer calculations to examine the multi-band polarization in the dust scattering model. We use the 3D Monte Carlo radiative transfer code developed by \citet[][]{Nagao2016}, Paper I, and \citet[][]{Nagao2018} (see Paper I for details). The similar calculations in Paper I are performed for SN light in the $U$, $B$, $V$, $R$, $I$, $J$, $H$ and $K_s$ bands. In this study, photons are emitted from the origin without taking into account the expansion of the SN photosphere, as in Paper I. The size of the SN photosphere is generally negligible (see Section 2.3 in Paper I). It should be noted that, in our study, re-emission from CS dust is ignored for simplicity, although both absorption and scattering are taken into account. This simplification does not affect results for optical light, because the unpolarized re-emitted light is mostly in near-infrared wavelengths. On the other hand, the calculated near-infrared polarization should be regarded as an upper limit in our calculations.

\subsection{Dust Model}
We adopt LMC-type, SMC-type and MW-type dust models (hereafter, the LMC, SMC and MW dust models, respectively). These dust models are constructed as a mixture of astronomical silicate and graphite grains of various sizes, assuming that dust grains have spherical shapes. It is assumed that the size distribution of dust grains is power-low with an index of $-3.5$, and that the maximum and minimum grain radii are $0.25$ $\mu$m and $0.005$ $\mu$m, respectively \citep[][]{Mathis1977}. The mass ratio between astronomical silicate and graphite is 0.843:0.157, 0.970:0.030 and 0.642:0.358 for the LMC, SMC and MW dust models, respectively. These values are determined so that the dust models well produce the observed extinction curves in the LMC, the SMC and the MW \citep[][]{Gordon2003, Whittet2003}. Additionally, we test a carbon dust model (hereafter, the C dust model). In this model, we adopt the same size distribution with the other models, except that the silicate fraction is set to be zero. The optical properties in these dust models are summarized in Table 1.

\begin{table*}
\caption{Dust parameters}
\begin{tabular}{|c|c|cccc|cccc|}
\tableline
&&\multicolumn{4}{|c|}{LMC} & \multicolumn{4}{|c|}{SMC}\\

\tableline
filter & $\lambda$[$\mu$ m] & $\kappa_{\mathrm{abs}}$[cm$^{2}$/g] & $\kappa_{\mathrm{scat}}$[cm$^{2}$/g] & $\kappa_{\mathrm{ext}}(\nu)/\kappa_{\mathrm{ext}}(U)$ & g & $\kappa_{\mathrm{abs}}$[cm$^{2}$/g] & $\kappa_{\mathrm{scat}}$[cm$^{2}$/g] & $\kappa_{\mathrm{ext}}(\nu)/\kappa_{\mathrm{ext}}(U)$ & g\\

\tableline \tableline
U & 0.36 & 1.206E+4 & 2.673E+4 & 1.000E+0 & 5.304E-1 & 5.534E+3 & 2.546E+4 & 1.000E+0 & 5.568E-1 \\ \hline
B & 0.44 & 9.744E+3 & 1.990E+4 & 7.641E-1 & 5.051E-1 & 4.281E+3 & 1.823E+4 & 7.265E-1 & 5.423E-1 \\ \hline
V & 0.55 & 7.735E+3 & 1.470E+4 & 5.783E-1 & 4.737E-1 & 3.316E+3 & 1.296E+4 & 5.251E-1 & 5.205E-1 \\ \hline
R & 0.66 & 6.172E+3 & 1.092E+4 & 4.406E-1 & 4.277E-1 & 2.561E+3 & 9.321E+3 & 3.834E-1 & 4.682E-1 \\ \hline
I & 0.81 & 4.739E+3 & 7.034E+3 & 3.035E-1 & 4.194E-1 & 2.035E+3 & 5.658E+3 & 2.482E-1 & 4.846E-1 \\ \hline
J & 1.2  & 2.704E+3 & 2.506E+3 & 1.343E-1 & 1.967E-1 & 1.126E+3 & 1.631E+3 & 8.894E-2 & 2.152E-1 \\ \hline
H & 1.7  & 1.713E+3 & 8.551E+2 & 6.621E-2 & 7.356E-2 & 7.359E+2 & 4.671E+2 & 3.882E-2 & 9.461E-2 \\ \hline
Ks & 2.2 & 1.162E+3 & 3.060E+2 & 3.785E-2 & 4.123E-2 & 5.414E+2 & 1.648E+2 & 2.279E-2 & 5.526E-2 \\ 
\tableline
\end{tabular}
\end{table*}

\setcounter{table}{0}
\begin{table*}
\caption{Dust parameters (continued)}
\begin{tabular}{|c|c|cccc|cccc|}
\tableline
&&\multicolumn{4}{|c|}{MW} & \multicolumn{4}{|c|}{C}\\

\tableline
filter & $\lambda$[$\mu$ m] & $\kappa_{\mathrm{abs}}$[cm$^{2}$/g] & $\kappa_{\mathrm{scat}}$[cm$^{2}$/g] & $\kappa_{\mathrm{ext}}(\nu)/\kappa_{\mathrm{ext}}(U)$ & g & $\kappa_{\mathrm{abs}}$[cm$^{2}$/g] & $\kappa_{\mathrm{scat}}$[cm$^{2}$/g] & $\kappa_{\mathrm{ext}}(\nu)/\kappa_{\mathrm{ext}}(U)$ & g\\

\tableline \tableline
U & 0.36 & 2.239E+4 & 2.875E+4 & 1.000E+0 & 4.934E-1 & 5.539E+4 & 3.518E+4 & 1.000E+1 & 4.038E-1 \\ \hline
B & 0.44 & 1.839E+4 & 2.253E+4 & 8.002E-1 & 4.573E-1 & 4.601E+4 & 3.095E+4 & 8.496E-1 & 3.592E-1 \\ \hline
V & 0.55 & 1.473E+4 & 1.745E+4 & 6.292E-1 & 4.187E-1 & 3.706E+4 & 2.625E+4 & 6.990E-1 & 3.203E-1 \\ \hline
R & 0.66 & 1.189E+4 & 1.345E+4 & 4.954E-1 & 3.834E-1 & 3.014E+4 & 2.153E+4 & 5.704E-1 & 3.115E-1 \\ \hline
I & 0.81 & 9.019E+3 & 9.213E+3 & 3.565E-1 & 3.559E-1 & 2.269E+4 & 1.617E+4 & 4.290E-1 & 2.678E-1 \\ \hline
J & 1.2  & 5.203E+3 & 3.892E+3 & 1.778E-1 & 1.843E-1 & 1.318E+4 & 8.317E+3 & 2.374E-1 & 1.725E-1 \\ \hline
H & 1.7  & 3.260E+3 & 1.469E+3 & 9.249E-2 & 6.296E-2 & 8.202E+3 & 3.431E+3 & 1.284E-1 & 5.452E-2 \\ \hline
Ks & 2.2 & 2.145E+3 & 5.294E+2 & 5.229E-2 & 3.431E-2 & 5.283E+3 & 1.243E+3 & 7.205E-2 & 2.887E-2 \\ 
\tableline
\end{tabular}
\end{table*}

\subsection{Distribution of CS Dust}
As is the case in Paper I, we adopt the blob configuration for the distribution of CS dust (see Figure 1). In this paper, the bipolar jet and disk configurations in Paper I are not considered. The bipolar jet model gives similar results with the blob model, and the disk model is unfavorable to explain the observed polarization feature in SNe IIP (see Paper I for details).

\begin{figure}[htbp]
  \includegraphics[scale=0.23]{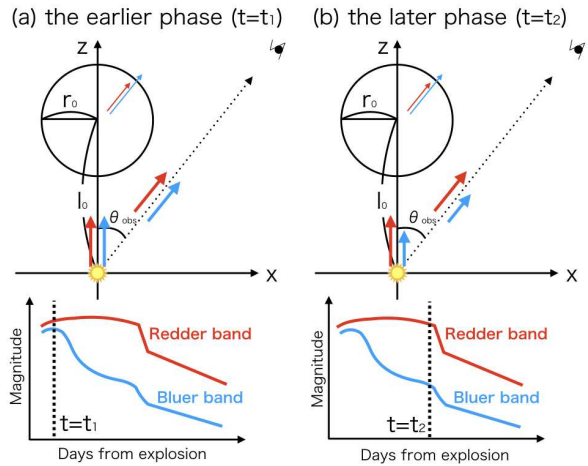}
  \caption{Geometry of the blob model and schematic picture of the wavelength dependence due to the different LC shapes and dust properties for different bands.}
\end{figure}

The distribution of CS dust is defined as in Paper I (see Figure 1). The distribution is specified by the distance of the blob from the SN and the radius of the blob ($l_0$ and $r_0$, respectively). The viewing angle is denoted as an angle $\theta_{\rm{obs}}$, which is the angle between the observer's direction and the direction to the blob. We adopt several values for $l_0$: 1.5, 2.0, 2.5, 3.0, 3.5, and 4.0$\times 10^{17}$ cm ($\sim 0.049$, $\sim 0.065$, $\sim 0.081$, $\sim 0.097$, $\sim 0.11$, and $\sim 0.13$ pc). The corresponding typical light travel time of the scattered light for each case is shown in Table 2, which is the same as Table 1 in Paper I. The value of $r_0$ is set so that a covering fraction of the blob for the SN light is 0.01 (i.e., the corresponding solid angle is $4 \pi \times 0.01$). Since the solid angle of the blob covering the SN light is $\pi r_{0}^{2}/l_{0}^{2}$, we set $r_0 = 0.2 l_0$. These $l_0$ and $r_0$ are the same as those in Paper I. It is assumed that the density of the blob ($\rho_0$) is uniform within it. We adopt the optical depth in the $U$ band along the diameter of the blob ($\tau_{0} (\rm{U})$) instead of $\rho_0$, as a model parameter: $\tau_{0} (\rm{U}) = 2r_0 \kappa_{\rm{ext},\nu}(\rm{U}) \rho_0 = 0.5, 1.0, 2.0, 3.0$, and $10.0$, where $\kappa_{\rm{ext},\nu}(\rm{U}) $ is the mass extinction coefficient in the $U$ band. In this model, we therefore leave $l_0$ and $\tau_{0} (\rm{U})$ as our tunable parameters.

\begin{table*}
\caption{The typical light travel time ([day]) in the blob models. This is the same table as table 1 in Paper I.}
\begin{tabular}{ccccccc}
\tableline
 $l_0 \times 10^{17}$ [cm] & light travel time [day] & $\theta_{\mathrm{obs}}=10$ deg & $\theta_{\mathrm{obs}}=30$ deg & $\theta_{\mathrm{obs}}=50$ deg & $\theta_{\mathrm{obs}}=70$ deg & $\theta_{\mathrm{obs}}=90$ deg\\

\tableline \tableline
1.5 & $\sim 58 (1-\cos \theta_{\mathrm{obs}})$  & $\sim 1$ & $\sim 8$  & $\sim 21$ & $\sim 38$  & $\sim 58$\\ \hline
2.0 & $\sim 77 (1-\cos \theta_{\mathrm{obs}})$  & $\sim 1$ & $\sim 10$ & $\sim 28$ & $\sim 51$  & $\sim 77$\\ \hline
2.5 & $\sim 97 (1-\cos \theta_{\mathrm{obs}})$  & $\sim 1$ & $\sim 13$ & $\sim 34$ & $\sim 64$  & $\sim 97$\\ \hline
3.0 & $\sim 116 (1-\cos \theta_{\mathrm{obs}})$ & $\sim 2$ & $\sim 16$ & $\sim 41$ & $\sim 76$  & $\sim 116$\\ \hline
3.5 & $\sim 135 (1-\cos \theta_{\mathrm{obs}})$ & $\sim 2$ & $\sim 18$ & $\sim 48$ & $\sim 89$  & $\sim 135$\\ \hline
4.0 & $\sim 154 (1-\cos \theta_{\mathrm{obs}})$ & $\sim 2$ & $\sim 21$ & $\sim 55$ & $\sim 102$ & $\sim 154$\\ \hline
\tableline
\end{tabular}
\end{table*}

In this model, the total dust mass $M_{\rm{dust}}$ and the corresponding mass-loss rate $\dot{M}_{\rm{gas}}$ for the CS dusty blob, are derived as in Paper I.
\begin{eqnarray}
M_{\rm{dust}} &=& \frac{4}{3}\pi r_{0}^{3} \rho_{0} = \frac{2\pi l_{0}^{2} \tau_0(U)}{75 \kappa_{\rm{ext}}(U)} \nonumber\\ 
&\sim& 2.0 \times 10^{-4} \Biggl( \frac{l_0}{3\times 10^{17}\; \rm{cm}} \Biggr)^{2} \Biggl( \frac{\tau_0(U)}{2.0} \Biggr) \nonumber\\
&&\Biggl( \frac{\kappa_{\rm{ext}}(U)}{3.879 \times 10^{4}\; \rm{cm}^{2}\; \rm{g}^{-1}} \Biggr)^{-1} \; \rm{M}_{\odot},\\
\dot{M}_{\rm{gas}} &\sim& \frac{M_{\rm{gas}}}{2 r_0 / v_{\rm{w}}} = \frac{\pi v_{\rm{w}} l_0 \tau_0(U)}{15 f_{\rm{dust}} \kappa_{\rm{ext}}(U)} \nonumber\\
&\sim& 5.2 \times 10^{-6} \Biggl( \frac{l_0}{3\times 10^{17}\; \rm{cm}} \Biggr) \Biggl( \frac{\tau_0(U)}{2.0} \Biggr) \nonumber\\
&&\Biggl( \frac{\kappa_{\rm{ext}}(U)}{3.879 \times 10^{4}\; \rm{cm}^{2}\; \rm{g}^{-1}} \Biggr)^{-1} \Biggl( \frac{f_{\rm{dust}}}{0.01} \Biggr)^{-1} \nonumber \\
&&\Biggl( \frac{v_{\rm{w}}}{10^{6}\; \rm{cm}\; \rm{s}^{-1}} \Biggr) \; \rm{M}_{\odot} \rm{yr}^{-1},
\end{eqnarray}
where $v_{\rm{w}} $ and $f_{\rm{dust}}$ are wind velocity of a progenitor star and a dust-to-gas ratio, respectively. Here, the value of $\kappa_{\rm{ext}}(U)$ is set to be that in the LMC dust model. For $f_{\rm{dust}}$ and $v_{\rm{w}}$, we adopt the typical values used in the literature \citep[e.g.,][]{Marshall2004, Mauron2011}, though the values are still observationally uncertain. As in Paper I, the range of mass-loss rate for the range of $l_0$ and $\tau_0(U)$ adopted in this paper ($6.4 \times 10^{-7} \lesssim \dot{M} \lesssim 3.4 \times 10^{-5} \; \rm{M}_{\odot} \rm{yr}^{-1}$) is consistent with those observationally derived for RSGs \citep[$1.0 \times 10^{-7} \lesssim \dot{M} \lesssim 1.0 \times 10^{-4} \; \rm{M}_{\odot} \rm{yr}^{-1}$, e.g.,][]{Mauron2011}.

\subsection{Input SN light curve model}
As an input SN light, we use two light curve (LC) models: the simple and realistic LC models.
In both LC models, the SN light is assumed to be unpolarized. The only source of polarization in this study is dust scattering. In this paper, we always use the Vega magnitude.

In the simple LC model, we assume the LCs in all the bands are the same as that in Paper I: The absolute magnitude of the SN is $-16$ mag up to $85$ days after the explosion (the plateau phase), $-13.5$ mag after $120$ days (the nebular phase) and the linearly interpolated value between the two phases ($85-120$ days). It should be noted that the important value to determine the polarization degree is not the absolute magnitude but the difference of the magnitude in the plateau and nebular phases (see Paper I).

In the realistic LC model, we adopt the multi-band LCs of SN 2004et provided by \citet[][]{Maguire2010} as typical ones for SNe IIP. This SN is one of the most well-observed SNe IIP and shows typical LC shapes. However it is unclear whether this $U$-band LC shape is typical one, just because of insufficiency of observations in the $U$ band due to observational difficulties. The other recent well-observed SNe IIP, e.g., SNe 2012aw and 2017eaw show the similar rapid decline in the $U$ band at early phase with SN 2004et, while the Nugent's template ({\url https://c3.lbl.gov/nugent/nugent\_templates.html}), which is based mainly on a model of SN 1999em, shows a slower decline in the $U$ band and a more similar evolution to the $B$ band. It should be noted that, in order to quantitatively compare the following results with observations, we have to calculate the polarization using the LC shapes of the SN we are interested in. We do not specify the absolute values of the SN magnitude. As shown in Paper I, the polarization degree in the dust scattering model does not depend on the absolute values of the SN magnitude (see also Section 3.2). The LCs of SN 2004et \citep[][]{Maguire2010} are shown in Figure 2. The missing data points are linearly interpolated for simplicity, and the values before the first observation are set to be the same value as the first observation for each band. This simplification does not change the results in this paper.

\begin{figure}[t]
  \includegraphics[scale=0.7]{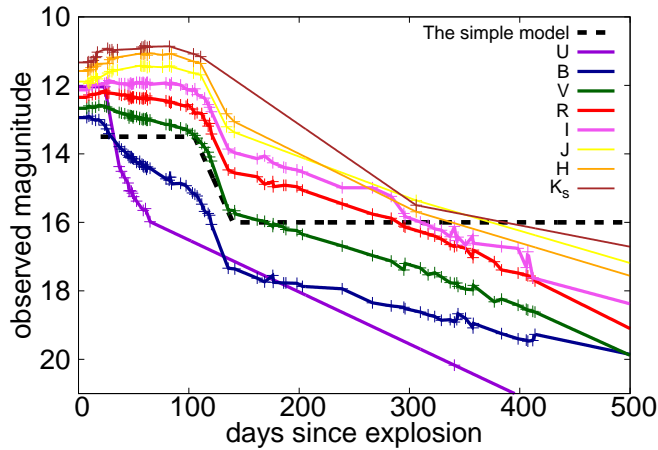}
  \caption{Light curves in SN 2004et (Maguire et al. 2010).}
\end{figure}

\section{Results}
We discuss the wavelength dependence of the polarization in the dust scattering model. Paper I suggested that the wavelength dependence of the polarization comes from the wavelength dependence of both dust optical parameters and input LC shapes toward a certain viewing direction: The dust optical parameters are represented by three parameters (the optical depth of CS dust, $\tau_0 (\nu)$, the albedo of CS dust, $\omega (\nu)$, and the scattering angle distribution of CS dust, $g(\nu)$), while the LC shapes are represented by the brightness changes of the input LCs ($\Delta M (\nu, t_{\rm{lt}})$) in the time scale of the delay time for the scattered photons (roughly the light travel time; $t_{\rm{lt}}$). In our calculations, we use the following parameters as tunable parameters: the input SN LC model (the simple or realistic LC models), the dust model (the LMC, SMC, MW or C dust models), the optical depth of the blob ($\tau_0(U)$), the distance of the blob from the SN ($l_0$) and the viewing angle of the observer ($\theta_{\rm{obs}}$). These tunable parameters determine the above wavelength dependence of the polarization; For example, the input LC is related with $\Delta M (\nu, t_{\rm{lt}})$. The dust model affects $\tau_0 (\nu)$, $\omega (\nu)$ and $g(\nu)$. The optical depth of the blob is connected with $\tau_0 (\nu)$. The distance and the viewing angle indirectly determine $\Delta M (\nu, t_{\rm{lt}})$ through the light travel time. In Section 3.1, we examine the wavelength dependence of the polarization due to the dust optical parameters ($\tau_0 (\nu)$, $\omega (\nu)$ and $g(\nu)$), by adopting the simple LC model. In Section 3.2, we discuss the wavelength dependence due to the LC shapes ($\Delta M (\nu, t_{\rm{lt}})$). Finally, we derive the wavelength dependence taking into account all these effects, in Section 3.3.

\subsection{Simple LC model}
In this subsection, we examine the wavelength dependence of the polarization due to the dust optical parameters ($\tau_0 (\nu)$, $\omega (\nu)$ and $g(\nu)$) toward various viewing directions ($\theta_{\rm{obs}}$), by calculating the polarization using the simple LC model. As long as the light travel time of the scattered echo is longer than $\sim 35$ days, the distance to the blob ($l_0$) does not affect the wavelength dependence in the case of the simple LC model (see Section 3.1.1 in Paper I for details). Therefore, the distance ($l_0$) is set to be $2.5 \times 10^{17}$ cm in this subsection. The remaining tunable parameters (the dust model, the optical depth of the blob ($\tau_0(U)$) and the viewing angle ($\theta_{\rm{obs}}$)) determine the wavelength dependence of the polarization through the wavelength dependence of the dust optical parameters ($\tau_0 (\nu)$, $\omega (\nu)$ and $g(\nu)$).

Figure 3 shows the time evolution of the polarization degree ($P$) for various bands, calculated toward different optical depth of the blob ($\tau_0(U)$) adopting the LMC dust model. The shape of the time-evolution curve of the polarization for each band can be interpreted by the discussions in Paper I. As in paper I, we call the maximum value and the full width at half maximum (FWHM) in each time-evolution curve of the polarization as $P_{\rm{max}}$ and $\Delta t$, respectively. The FWHM of the polarization evolution ($\Delta t$) is mainly determined by the light traveling time for the adopted tunable parameters, though the optical depth of CS dust ($\tau_0(\nu)$) also affects the FWHM ($\Delta t$) through multiple scattering processes (see Section 3 in Paper I for details). Thus, the FWHM ($\Delta t$) is almost constant for different bands, in the case of $\tau_0(U)=1.0$. In the case of $\tau_0(U)=10.0$, the FWHM ($\Delta t$) is smaller for shorter wavelengths. This is because, in the case with a high $\tau_0(\nu)$ (in other words, with a short wavelength), photons from the far side of the blob toward the observer, which contribute to the polarization at the later phase, are selectively absorbed by the dust in the blob. 

\begin{figure*}[t]
  \includegraphics[scale=0.7]{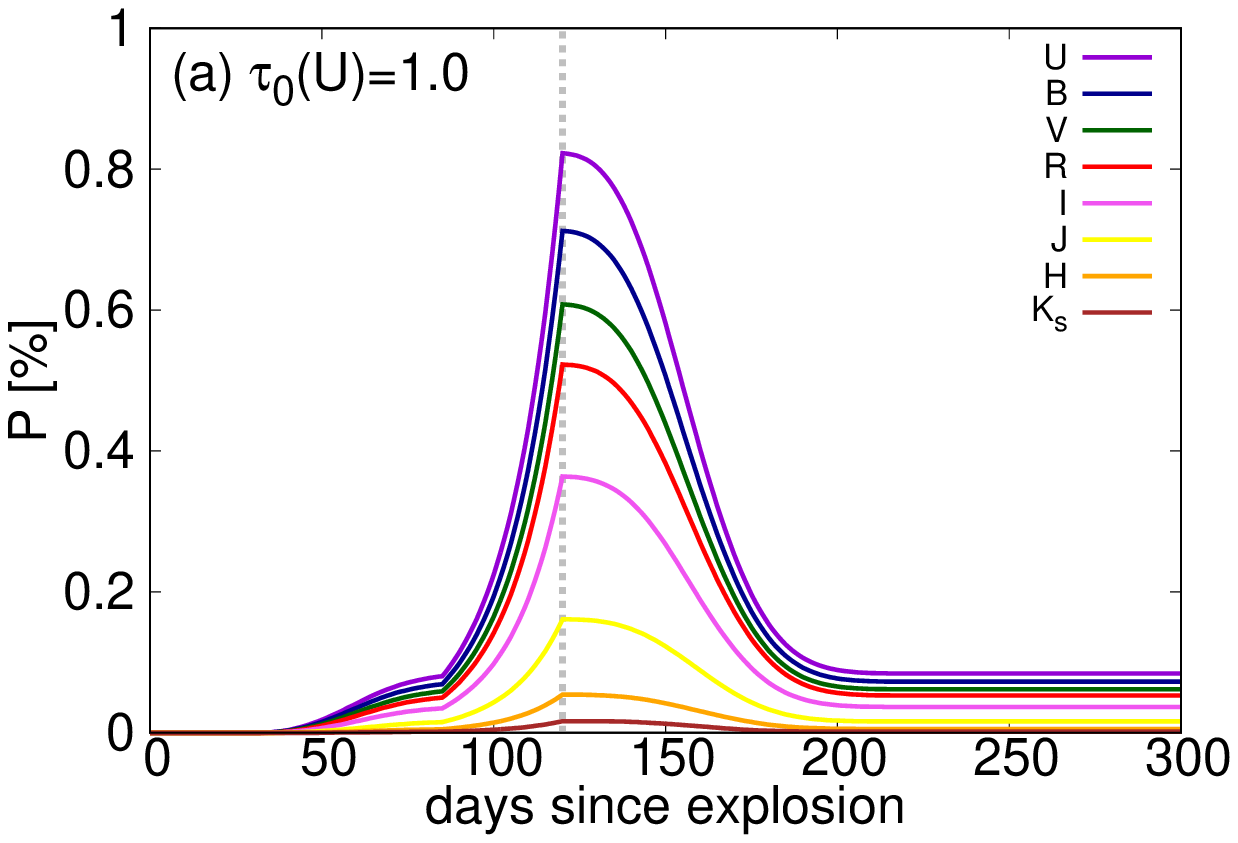}
  \includegraphics[scale=0.7]{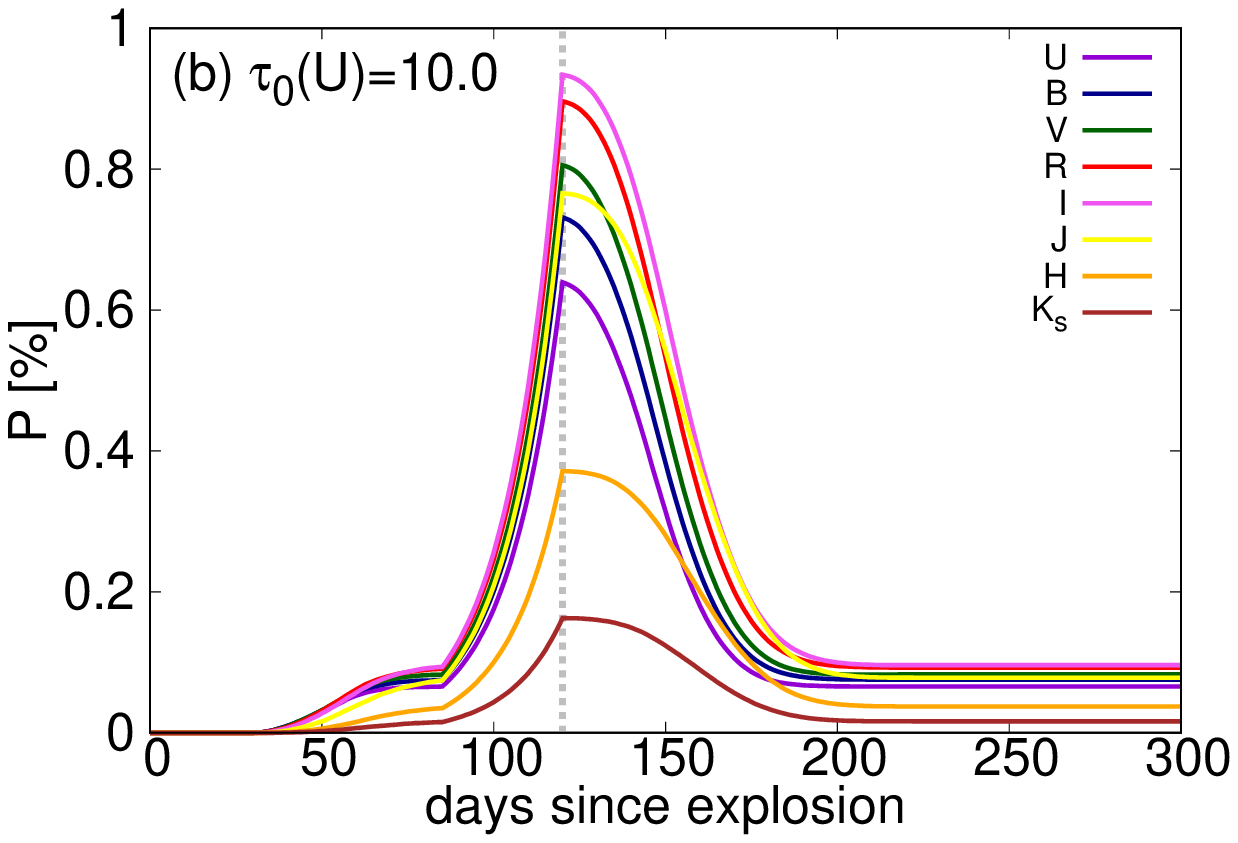}
  \caption{(a) Time evolution of the polarization in the LMC dust model, where $\tau_{0}(U) = 1.0$, $l_{0}=2.5 \times 10^{17}$ cm and $\theta_{\rm{obs}}=70^\circ$. (b) Same as (a), but for $\tau_{0}(U) = 10.0$. The vertical dotted lines show the epoch when the $B$-band polarization is maximized (i.e., 120 days, see Section 2.3.1).}
\end{figure*}

On the other hand, the maximum polarization degree ($P_{\rm{max}}$) is largely different for the different bands, which come from the wavelength dependence of the dust optical parameters ($\tau_0 (\nu)$, $\omega (\nu)$ and $g(\nu)$). Figure 4 shows the wavelength dependence of the polarization when the polarization degree in the $U$ band is maximized (For example, the timing shown in Figure 3 with dashed lines). In the case of $\tau_0(U)=1.0$, where the optical depth for all the bands is less than unity, the polarization degree in a shorter wavelength is larger. While, the polarization degree is maximized around the $R$ or $I$ band in the case of $\tau_0(U)=10.0$. These behaviors are interpreted mostly as effects of multiple scattering, which lead to depolarization of light (see Paper I). In the case with $\tau_0$ higher than $\sim 2$, the polarization does not become higher due to multiple scattering, even though the flux of the scattered echo is higher.

\begin{figure*}[t]
  \includegraphics[scale=0.7]{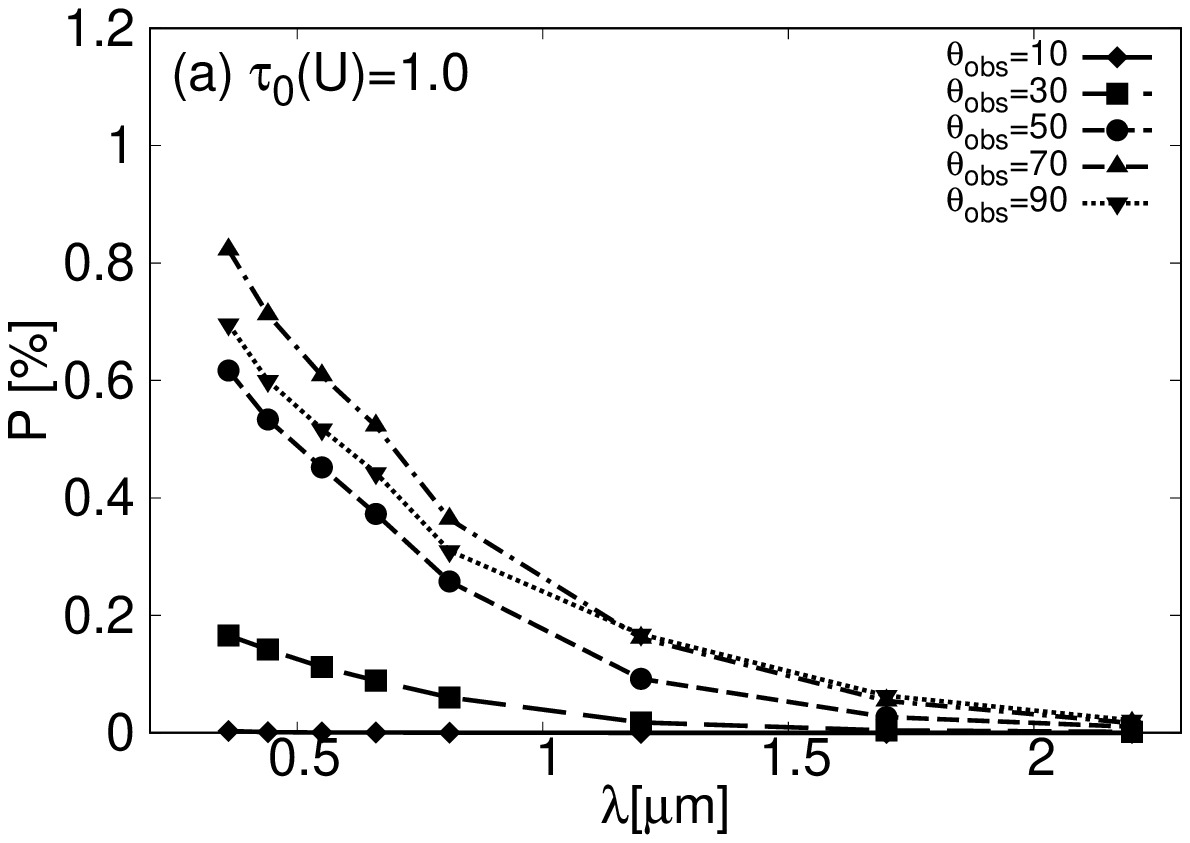}
  \includegraphics[scale=0.7]{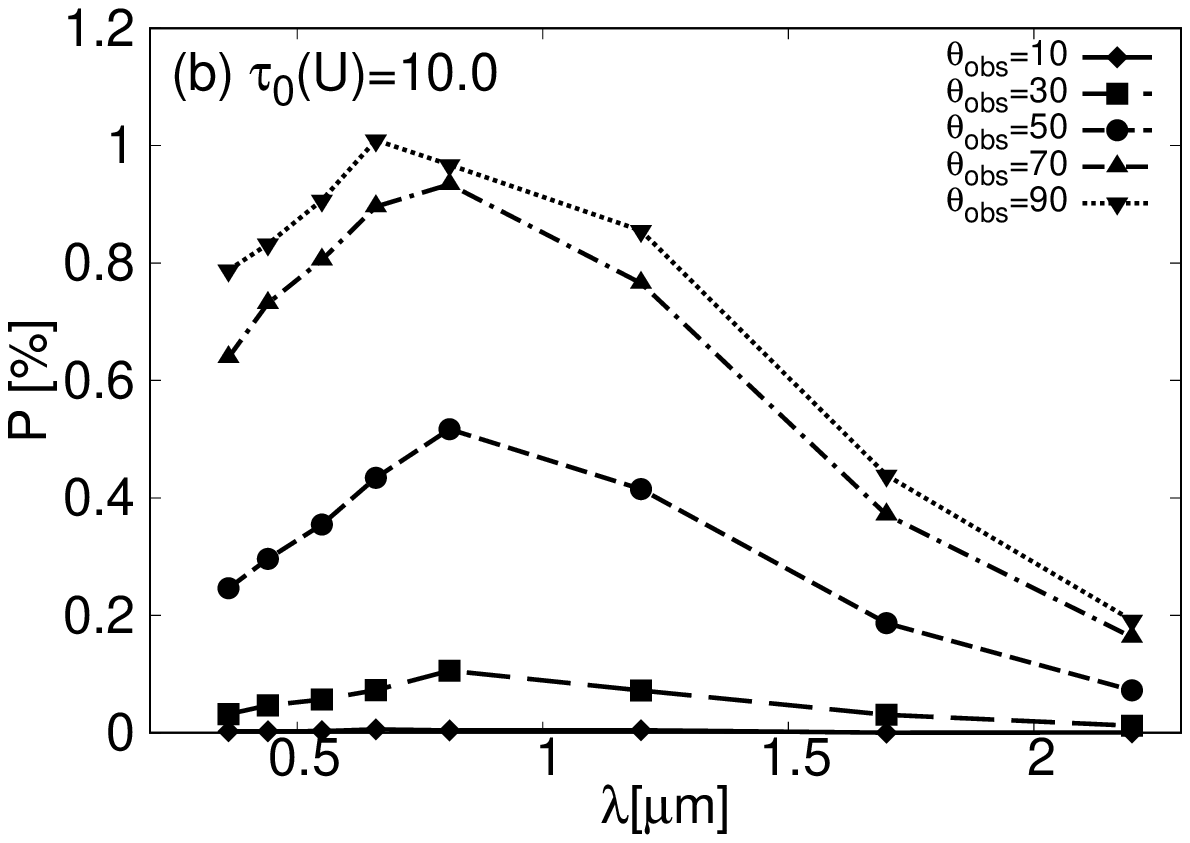}
  \caption{(a) Wavelength dependence of the polarization in the LMC dust model for different values of $\theta_{\rm{obs}}$, where $t=120$ days, $\tau_{0}(U) = 1.0$ and $l_{0}=2.5 \times 10^{17}$ cm. (b) Same as (a), but for $\tau_{0}(U) = 10.0$.}
\end{figure*}

To distinguish the effects of the albedo ($\omega (\nu)$) and the scattering angle distribution ($g(\nu)$) from those of the optical depth of CS dust($\tau_0 (\nu)$), we multiply the polarization degree in Figure 4(a) by ($\tau_{0}(\nu)/\tau_{0}(U))^{-1}(=\kappa_{\rm{ext}}(U)/\kappa_{\rm{ext}}(\nu)$). We can now see the wavelength dependence only from the albedo ($\omega (\nu)$) and the scattering angle distribution ($g(\nu)$), which is shown in Figure 5. This is because the polarization degree is proportional to the optical depth ($\tau_0(\nu)$) as long as $\tau_0(\nu) \lesssim 2$ (see Paper I). For longer wavelengths, the scattering angle distribution is more isotropic, which leads the scattered echo for larger $\theta_{\rm{obs}}$ to be higher. At the same time, the albedo ($\omega (\nu)$) become lower for longer wavelengths. Thus, the effects on the wavelength dependence from the albedo ($\omega (\nu)$) and the scattering angle distribution ($g(\nu)$) are cancelled out, and the wavelength dependence after all is not large (see Figure 5). The wavelength dependence of the optical depth of CS dust ($\tau_0(\nu)$) is the most important, if the difference of the input LCs for each band would be ignored. 

\begin{figure}[t]
  \includegraphics[scale=0.7]{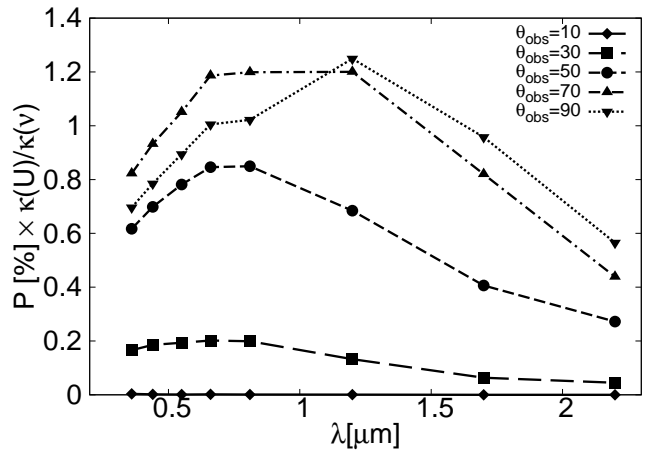}
  \caption{Wavelength dependence of the polarization from $\omega (\nu)$ and $g(\nu)$.
This is the same as Figure 4a, but the values have been multiplied by $\kappa_{\rm{ext}}(U)/\kappa_{\rm{ext}}(\nu)$ (the LMC dust model, $\tau_{0}(U) = 1.0$ and $l_{0}=2.5 \times 10^{17}$ cm).}
\end{figure}

Figure 6 shows the wavelength dependence of the polarization for various dust models. In the case of $\tau_0(U)=1.0$, the dependence become steeper in the order of the C, MW, LMC, SMC dust models, which is the ascending order of silicate in their dust composition. In the case of $\tau_0(U)=10.0$, the peak wavelength, which is correspond to the wavelength such as $\tau_0(\nu) \sim 2$, is different for different dust models. The both features are roughly interpreted as the difference of the wavelength dependence of $\kappa_{\rm{ext}}(\nu)$ for different dust models.

\begin{figure*}[t]
  \includegraphics[scale=0.7]{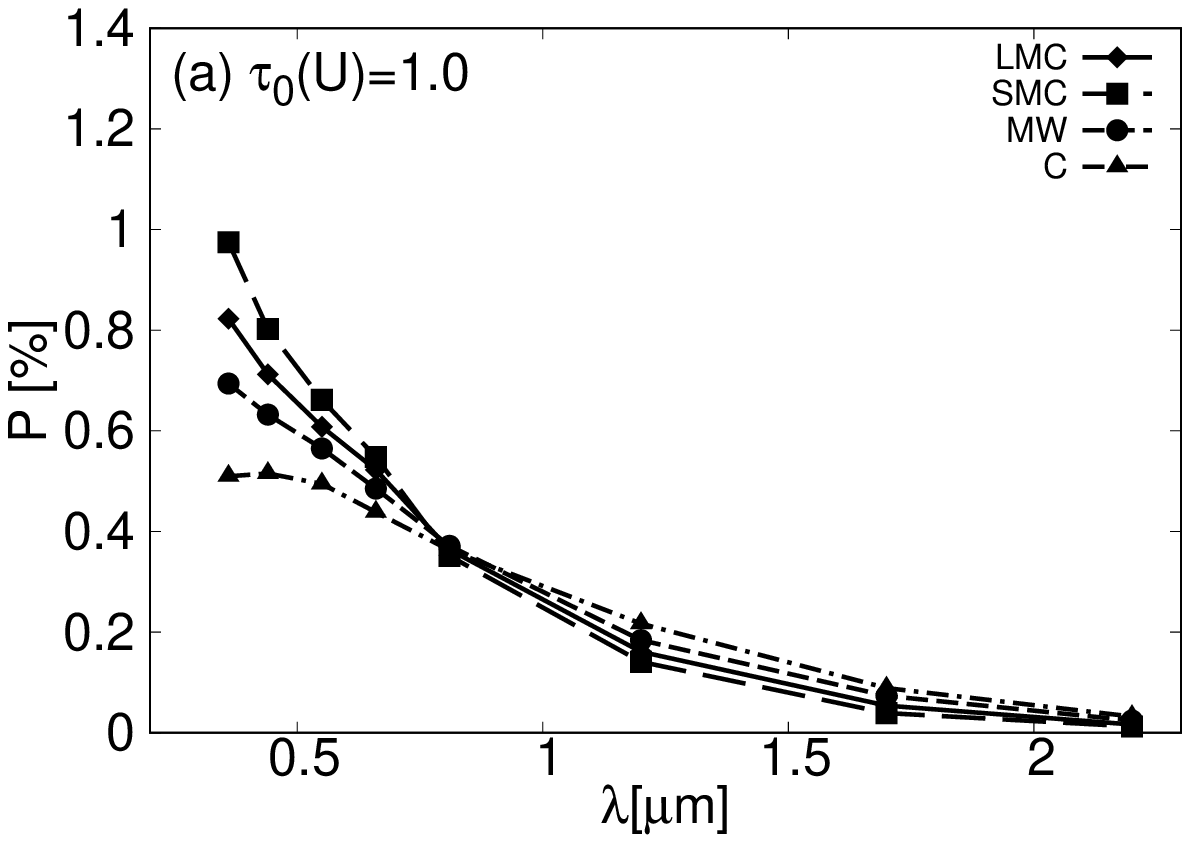}
  \includegraphics[scale=0.7]{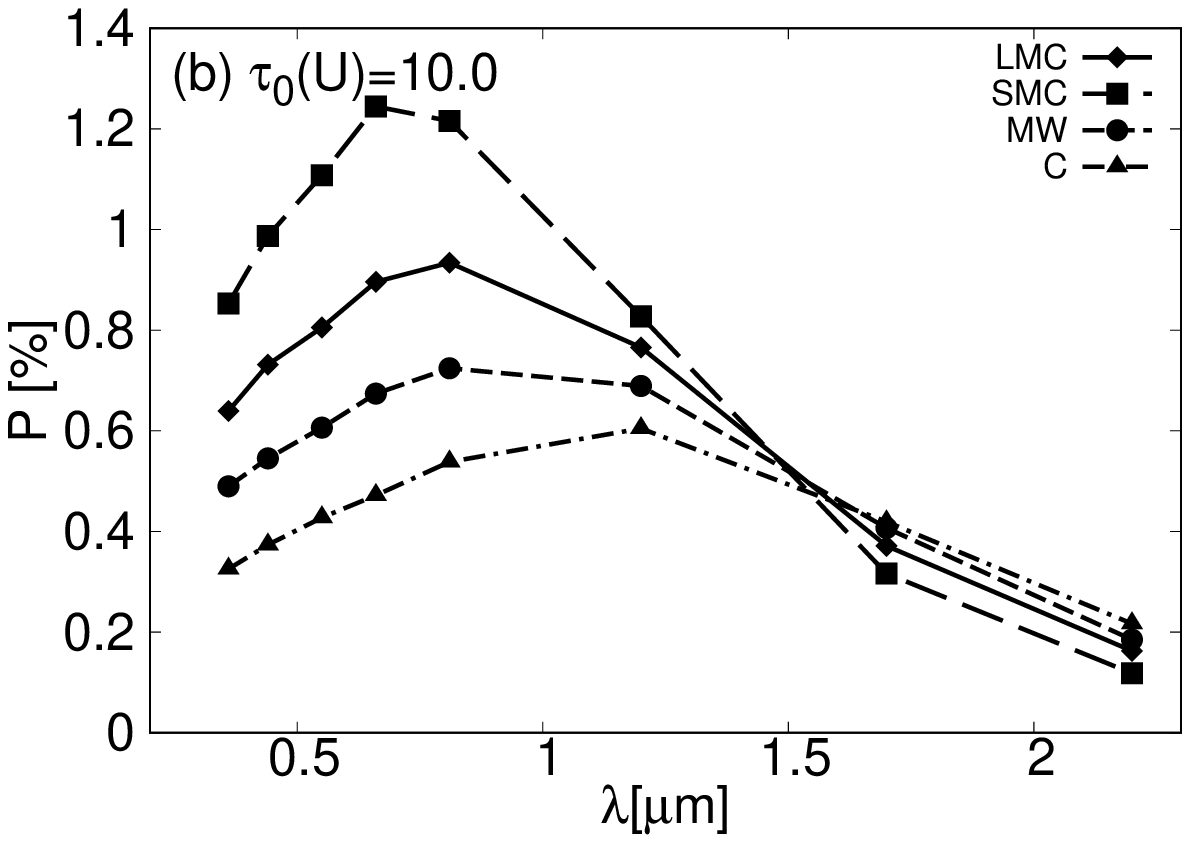}
  \caption{Same as Figure 4, but for various dust models with $\theta_{\rm{obs}}=70^\circ$ ($l_{0}=2.5 \times 10^{17}$ cm).}
\end{figure*}

\subsection{Realistic LC model}
In this subsection, we discuss the wavelength dependence of the polarization taking into account all the effects from the quantities: The dust optical parameters ($\tau_0 (\nu)$, $\omega (\nu)$, $g(\nu)$) and the LC shapes ($\Delta M (\nu, t_{\rm{lt}})$). We calculate the polarization using the realistic SN LC model as an input SN light. 

As discussed in Paper I, the polarization degree is determined by the flux ratio of the polarized scattered echo to the unpolarized intrinsic SN light. The flux of the scattered echo at time $t$ is proportional to the SN flux the light travel time ($t_{\rm{lt}}$) before the time $t$. Therefore, the ratio of the SN flux at $t-t_{\rm{lt}}$ to that at $t$ is a useful indicator to understand the wavelength dependence of the polarization due to the difference of input LC shapes in different bands. Figure 7 shows this flux ratio for various $t_{\rm{lt}}$. The general behavior for each band except the $U$ and $K$ bands is roughly the same with the case of the simple LC model. In the cases with lager $t_{\rm{lt}}$, the deviation from the value in the simple LC model is larger, due to the simplification in the simple LC model about the time evolution of the SN flux in the plateau and nebular phases. Therefore, the wavelength dependence is not large especially between the $B$ and $V$ bands and between the $R$, $I$, $J$ and $H$ bands, as long as the light travel time is not so long. In the case with longer light travel time, the polarization in the shorter wavelengths is enhanced because of the steeper decline in the plateau phase.

\begin{figure*}[t]
  \includegraphics[scale=0.7]{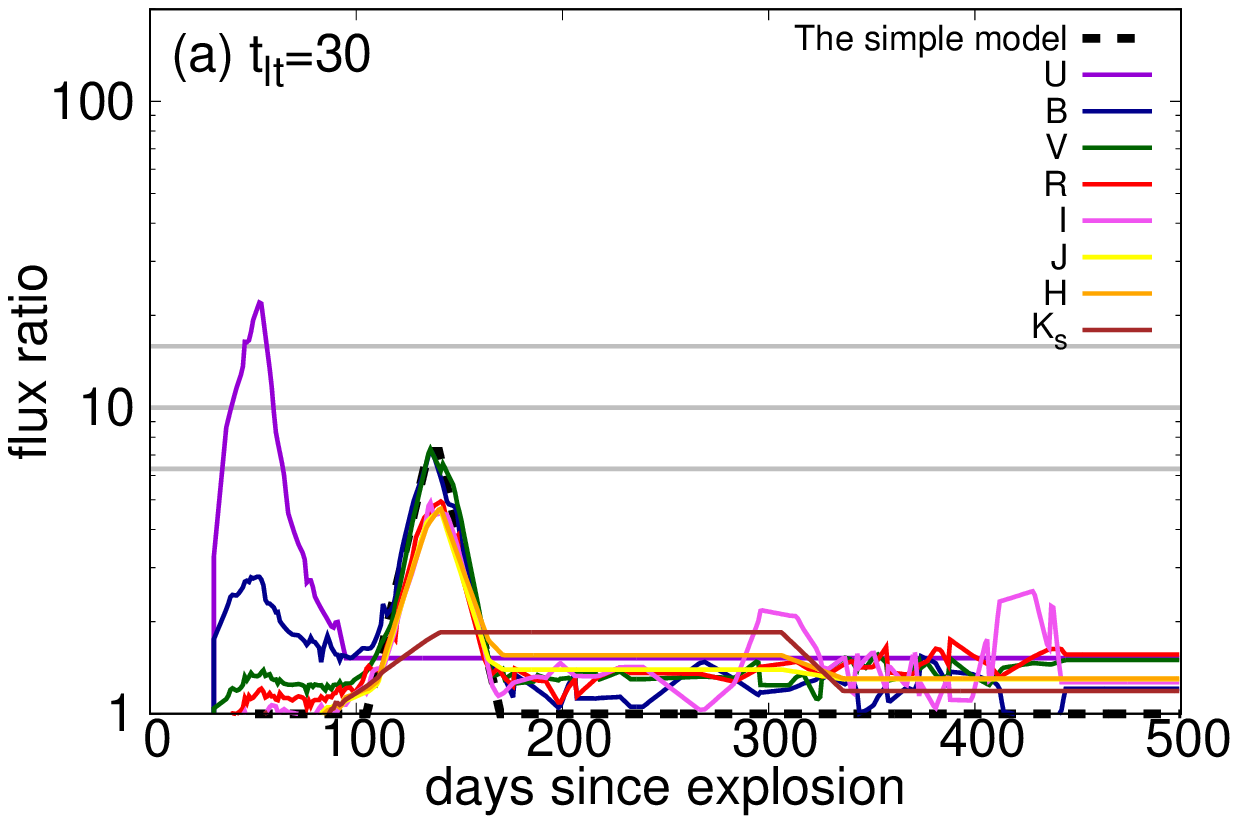}
  \includegraphics[scale=0.7]{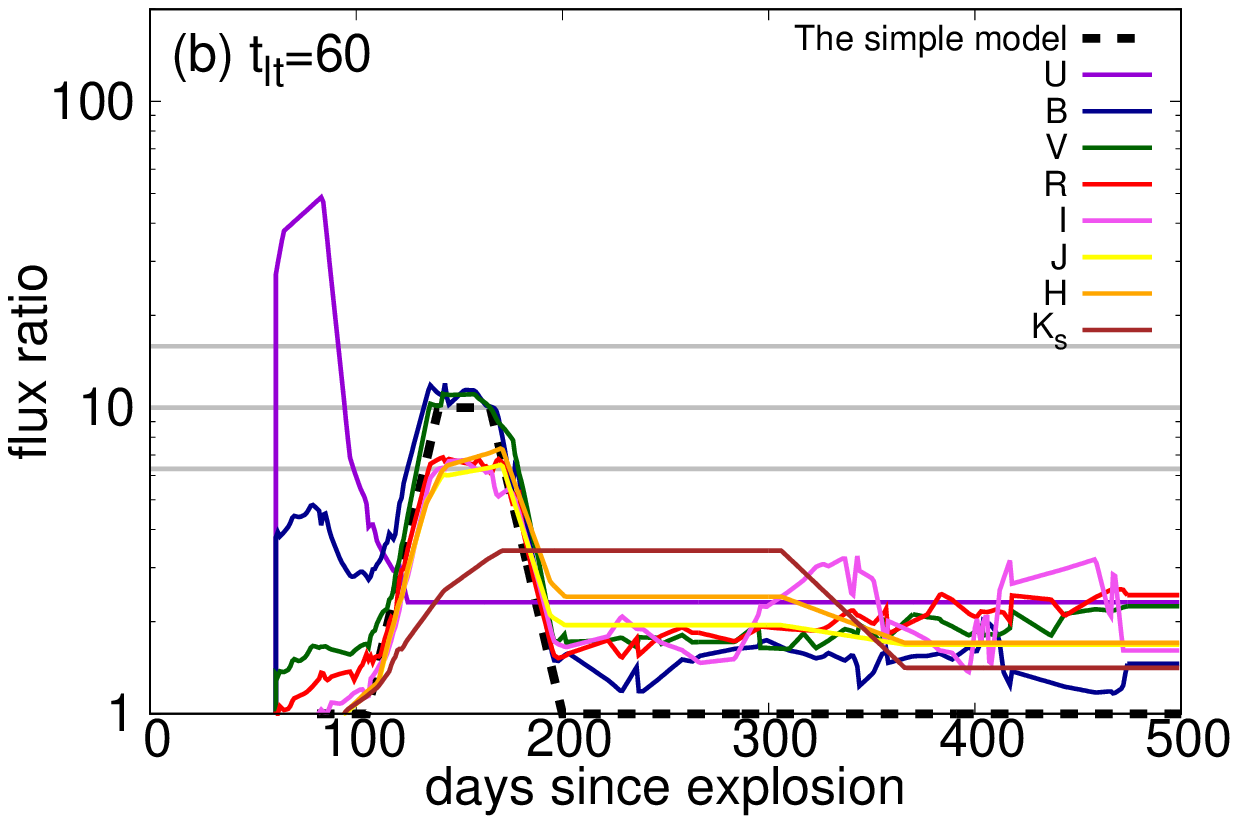}
  \includegraphics[scale=0.7]{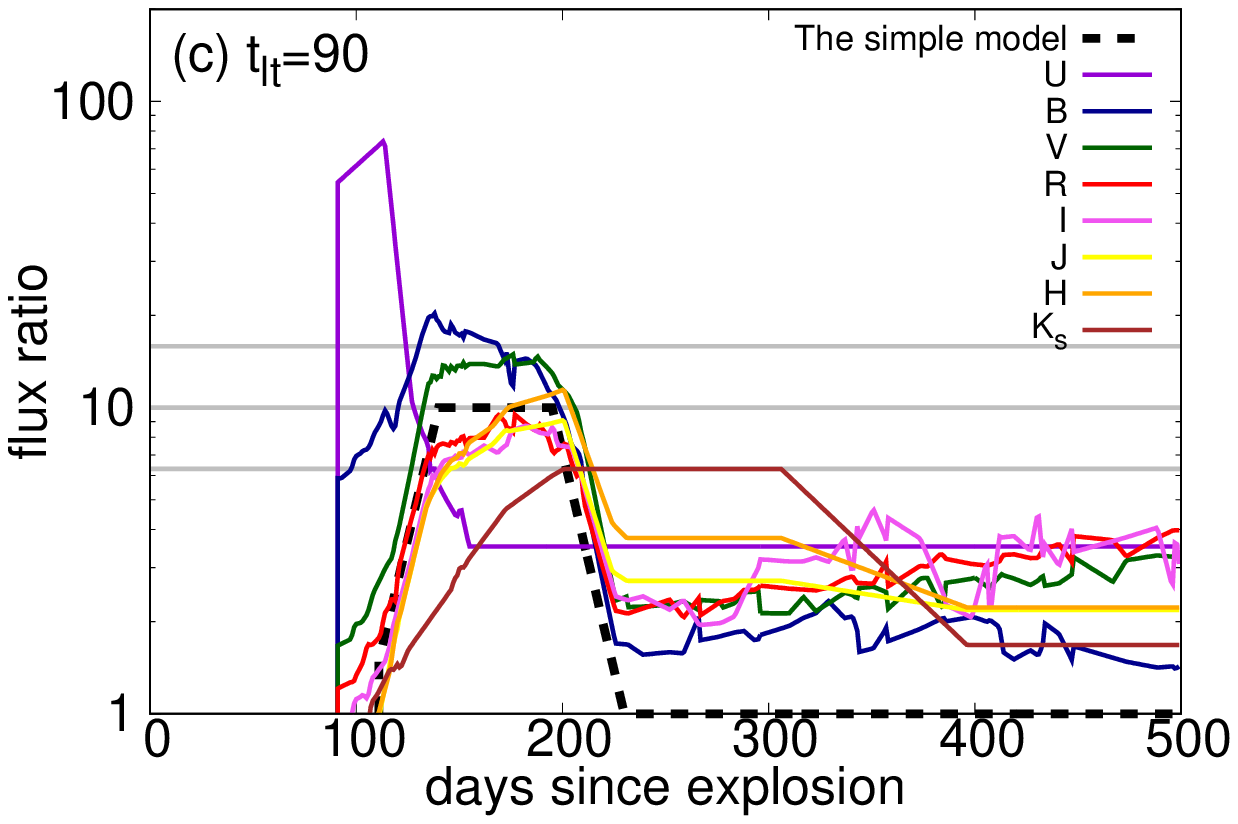}
  \includegraphics[scale=0.7]{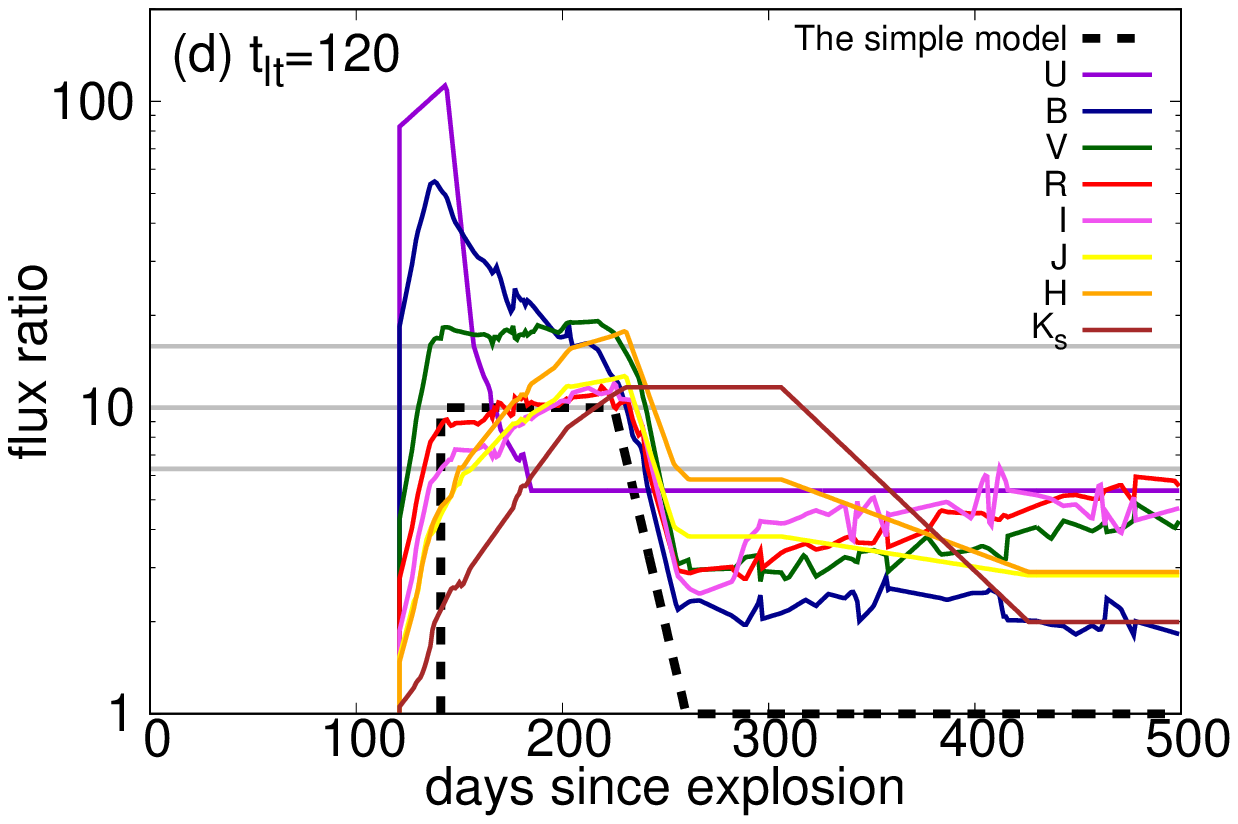}
  \includegraphics[scale=0.7]{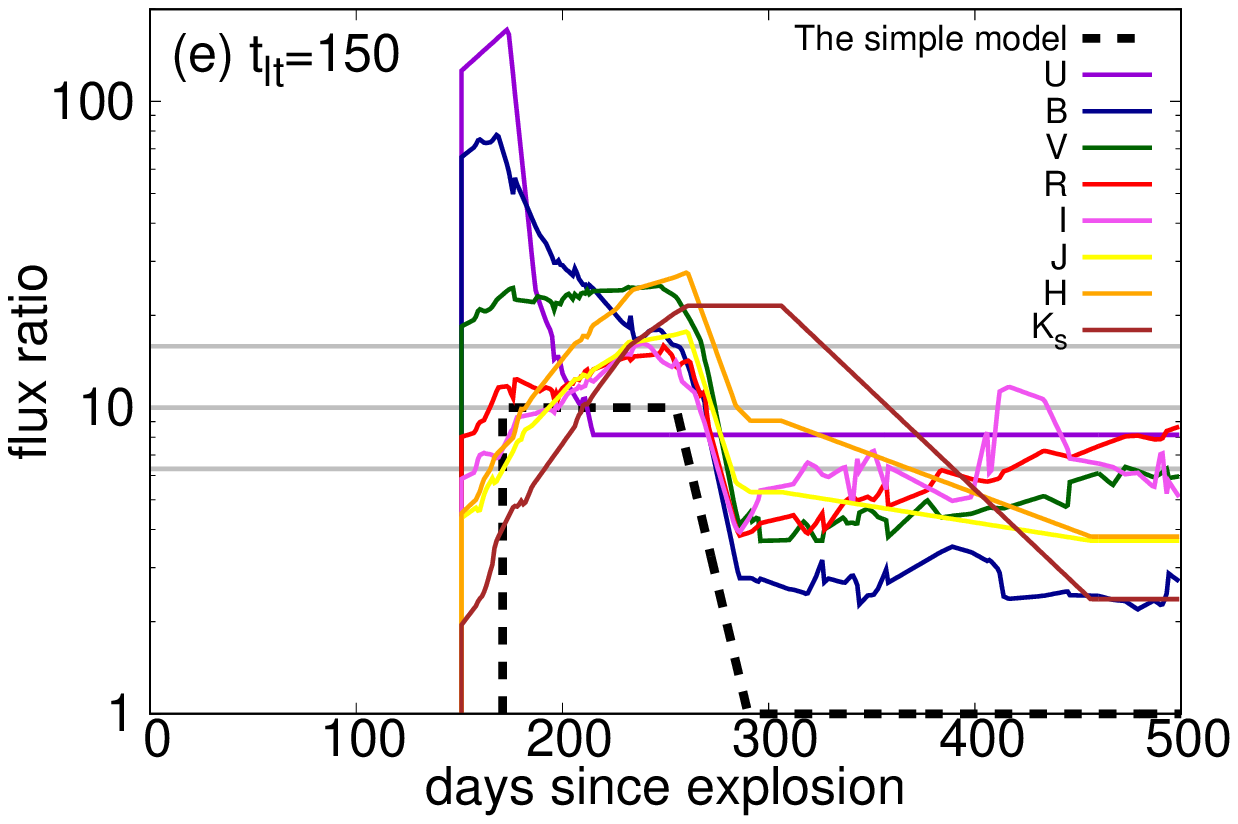}
  \caption{Flux ratio of the flux in SN 2004et at $t=t-t_{\rm{lt}}$ to that at $t$ for various values of $t_{\rm{lt}}$ ([days]). The horizontal lines express the corresponding flux ratio to the difference in magnitudes, $\Delta M =$ 2.0, 2.5 and 3.0 mag in the simple LC model, respectively.}
\end{figure*}

The time-evolution of the flux ratio in the $U$ band is different from those in the other bands. This is because the $U$-band LC that we adopt here shows a rapid decline since the early phase unlike in the other bands (see Figure 2). The information on the time and polarization degree when the $U$-band polarization is maximized is therefore helpful to estimate the parameters in our CS dust model. The time corresponds to the light travel time, which is related to the distance to the blob ($l_0$) and the viewing angle ($\theta_{\rm{obs}}$). The degree is connected mainly with the optical depth of the blob ($\tau_0(U)$) (and weakly with the dust model). As for the $K$ band, the LC we adopted is not good enough to derive strong conclusions: There is no observation during the transition from the plateau phase to the nebular phase, which is an important phase to create the polarization. Moreover, the polarization degree in the $K$ band is typically low, due to the unpolarized re-emission from the CS dust. Therefore, we do not discuss the $K$ band further in this paper.

Figure 8 shows the time evolution of the polarization degree $P$ for various bands, calculated toward different optical depth of the blob ($\tau_0(U)$) adopting the LMC dust model. This figure is the same as Figure 3, but for the realistic LC model instead of the simple LC model. As discussed {\bf above}, the $U$-band polarization shows different evolution from those in the other bands, due to the difference in the LC shapes. As is mentioned above, the information on the time and polarization degree of the $U$-band polarization at its peak are important to constrain the optical depth ($\tau_0(U)$), the distance to the blob ($l_0$) and the viewing direction ($\theta_{\rm{obs}}$). This feature in the $U$ band (the high polarization degree and the earlier emergence of the polarization) is one of the strong predictions from the dust scattering model. Roughly speaking, the polarization degree in the other bands in Figure 8 increases at the transition from the plateau phase to nebular phase, and then decreases in the light travel time. The FWHM of the polarization evolution ($\Delta t$) can be interpreted as in Figure 3: $\Delta t$ is relatively the same for each band in the case with $\tau_0(U)=1.0$, while $\Delta t$ is smaller for the shorter wavelengths in the case with $\tau_0(U)=10.0$.

\begin{figure*}[ht]
  \includegraphics[scale=0.7]{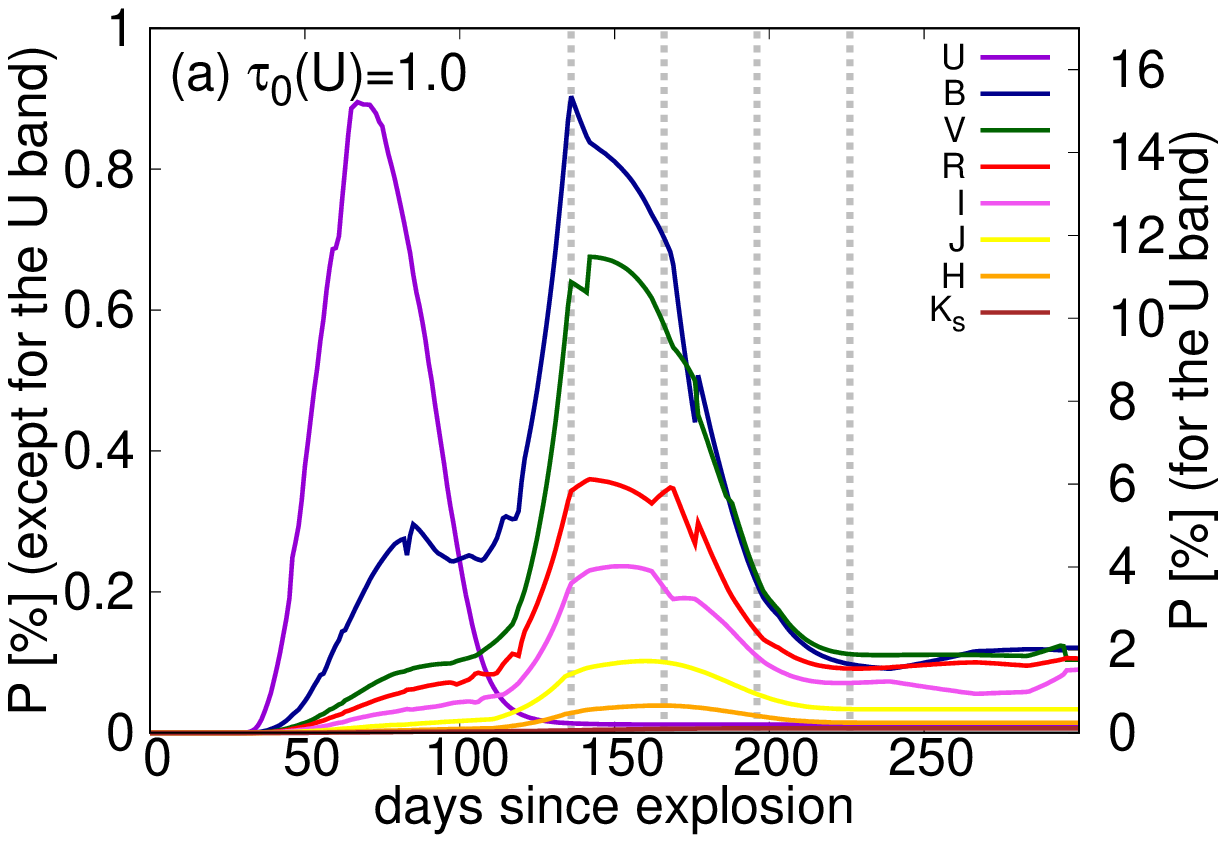}
  \includegraphics[scale=0.7]{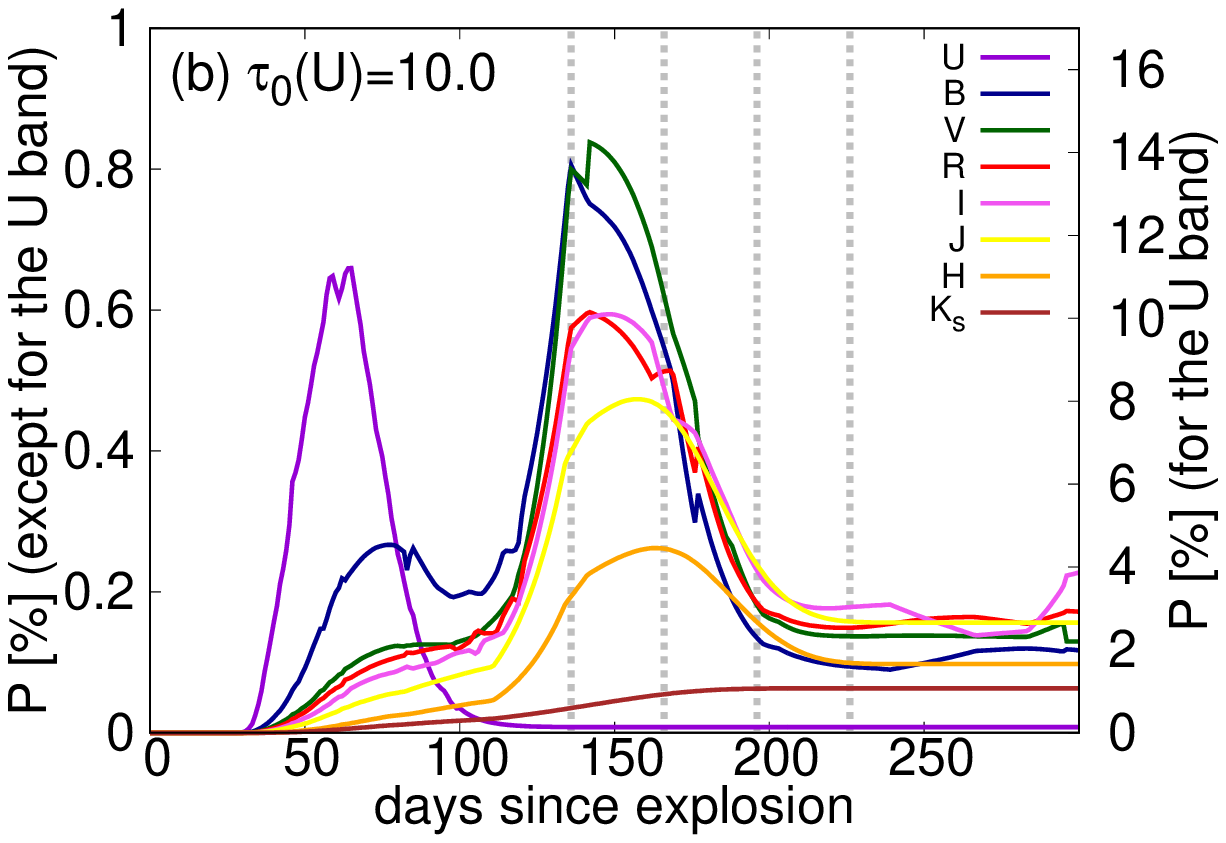}
  \caption{Same as Figure 3, but for the realistic LC model (the LMC dust model, $l_{0}=2.5 \times 10^{17}$ cm and $\theta_{\rm{obs}}=70^\circ$). The vertical dotted lines show the epoch $t=136, 166, 196$ and $226$ days, for Figure 9.}
\end{figure*}

Next, we discuss the wavelength dependence of the polarization in Figure 8. Figure 9 shows time evolution of the wavelength dependence of the polarization in Figure 8. The wavelength dependence evolves with time, which originates from the difference of the LC shapes for various bands. Hereafter, we do not consider the time evolution of the polarization, but focus on the polarization when the polarization degree is maximized in the $B$ band. This is roughly the timing when the polarization in all the bands, except for the $U$ and $K$ bands, is maximized. In the case with $\Delta t \lesssim 90$ days, the $U$-band polarization decreases until this time, while the $U$-band polarization is still surviving at this time if $\Delta t \gtrsim 90$ days (see Figure 7).

\begin{figure*}[ht]
  \includegraphics[scale=0.7]{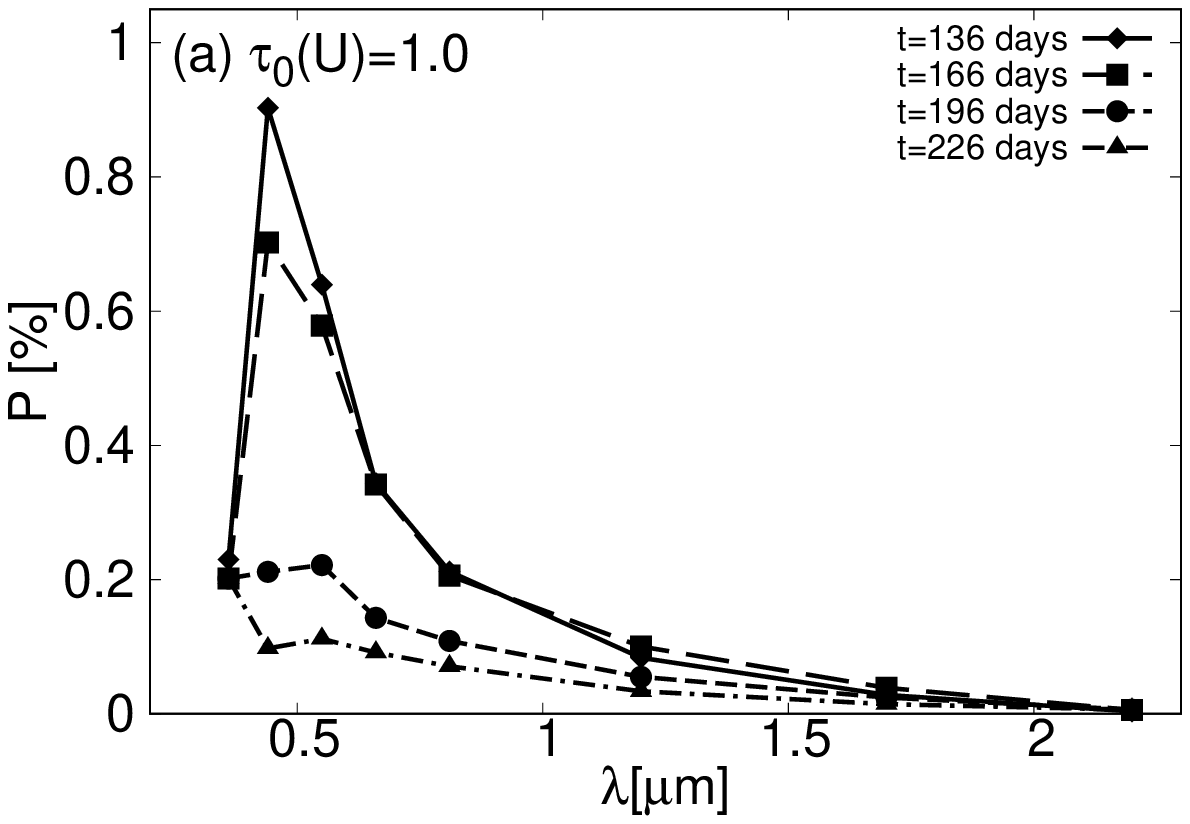}
  \includegraphics[scale=0.7]{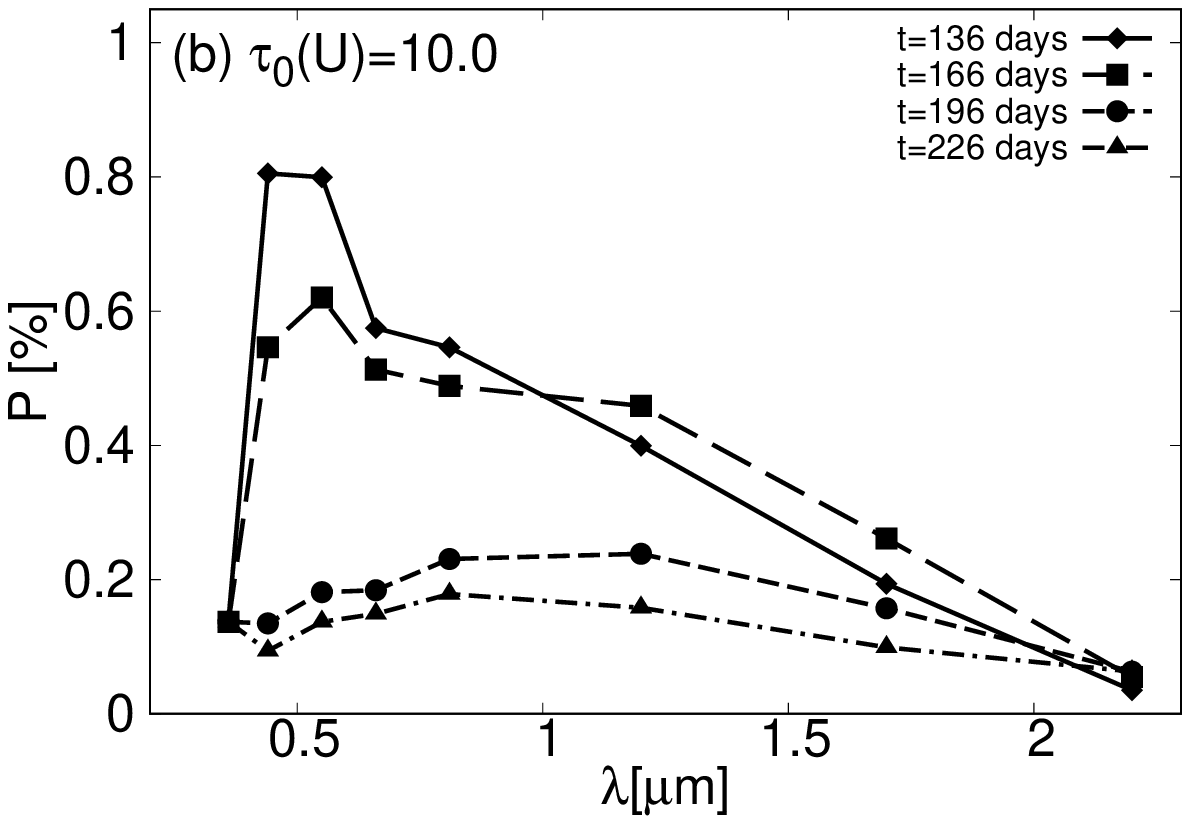}
  \caption{Time evolution of the wavelength dependence of the polarization in Figure 8 for $t=136, 166, 196$ and $226$ days (the LMC dust model, $l_{0}=2.5 \times 10^{17}$ cm and $\theta_{\rm{obs}}=70^\circ$).}
\end{figure*}

Figure 10 shows the wavelength dependence of the polarization toward various viewing directions when the $B$-band polarization is maximized. In the case of $\tau_0(U)=1.0$, where the optical depth for all the bands is less than unity, the polarization degree for shorter wavelength is higher, except that in the $U$ band (see above for the $U$-band polarization). This feature become stronger for larger viewing direction ($\theta_{\rm{obs}}$), because the light travel time is longer and the flux ratio of the SN and the echo is higher (see above). This feature is due to the effects of different LC shapes for different bands, which is not seen in Figure 4.  In the case of $\tau_0(U)=10.0$, the wavelength dependence become weaker than that in the case of $\tau_0(U)=1.0$, because the polarization do not become higher anymore due to multiple scatterings for the shorter wavelength.

\begin{figure*}[h]
  \includegraphics[scale=0.7]{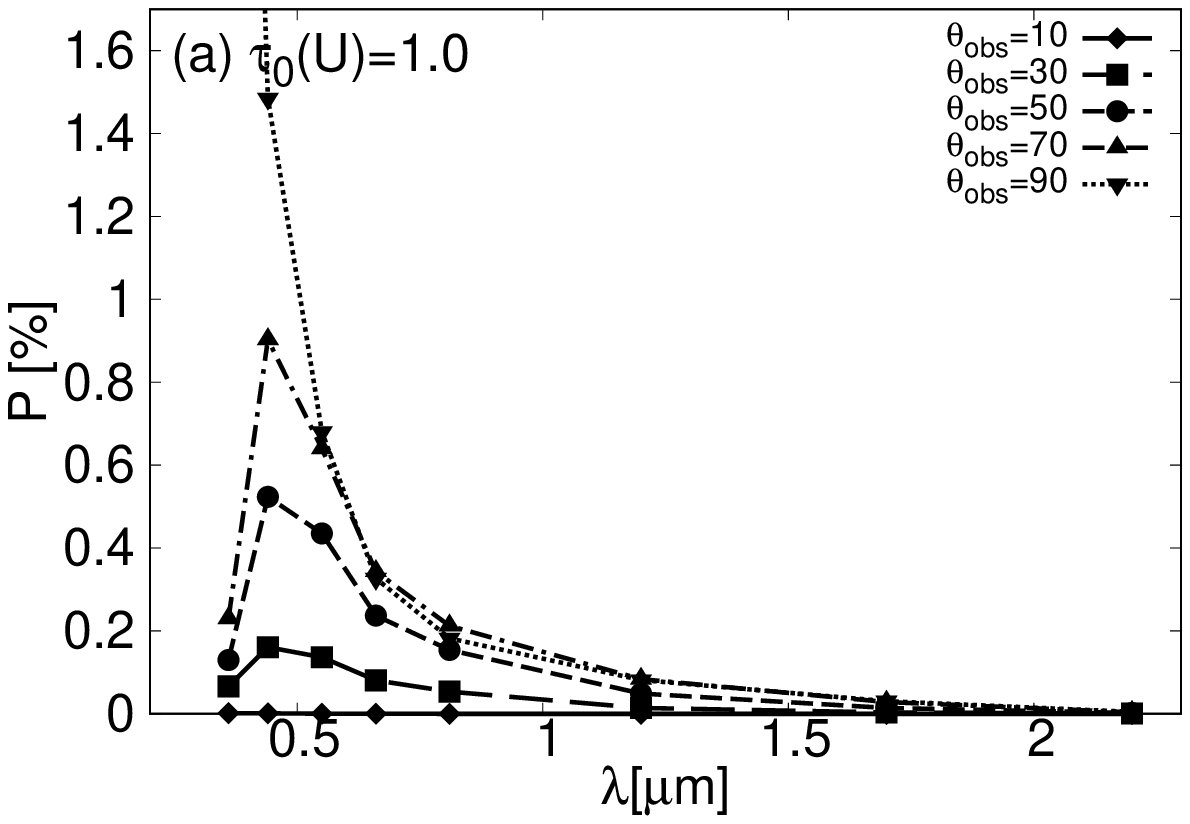}
  \includegraphics[scale=0.7]{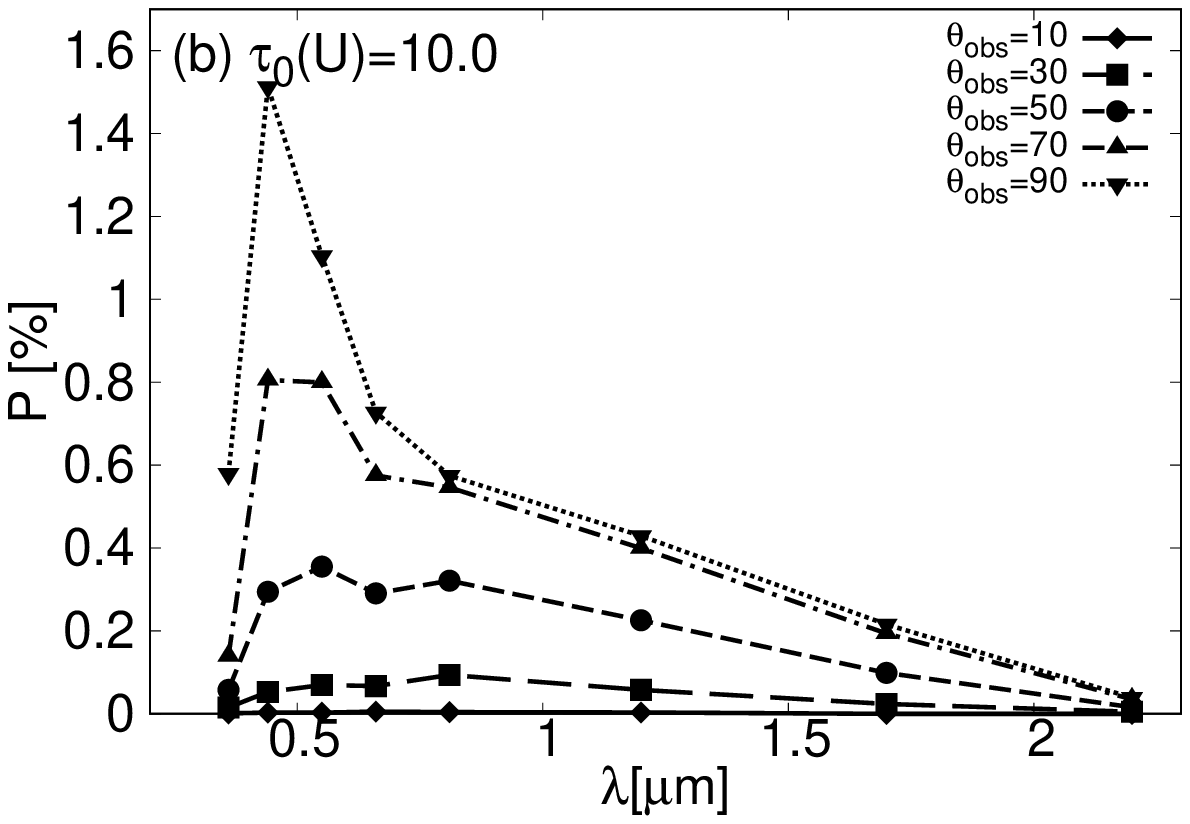}
  \caption{Same as Figure 4, but for the realistic LC model (the LMC dust model and $l_{0}=2.5 \times 10^{17}$ cm).}
\end{figure*}

Unlike the case for the simple SN LCs, the wavelength dependence depends on the distance to the blob ($l_0$), which is related with the light travel time ($t_{\rm{lt}}$) together with the viewing direction ($\theta_{\rm{obs}}$). Figure 11 is the wavelength dependence of the polarization for various distance of the blob ($l_0$). The wavelength dependence becomes stronger for larger distance ($l_0$). This is because the flux ratio of the SN and the echo in the shorter wavelength become higher as the light travel time becomes longer (see Figure 7 and above). 

\begin{figure*}[h]
  \includegraphics[scale=0.7]{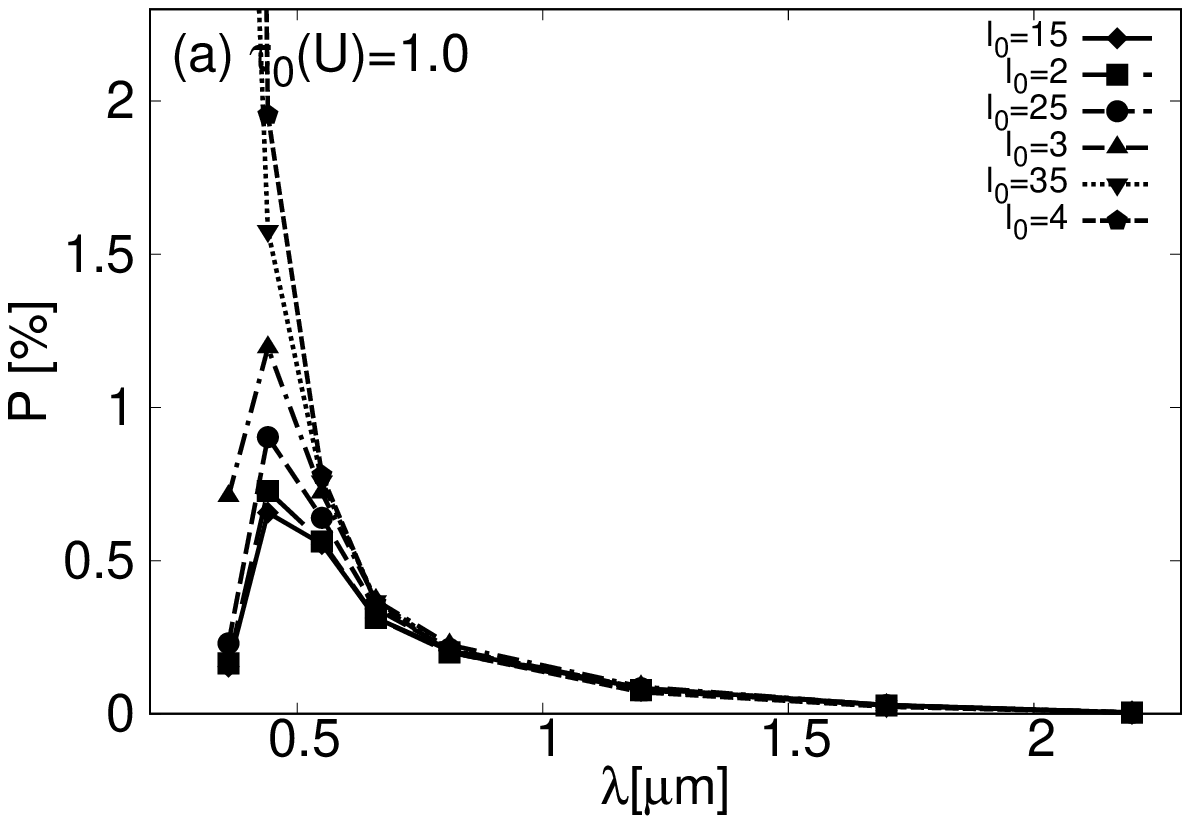}
  \includegraphics[scale=0.7]{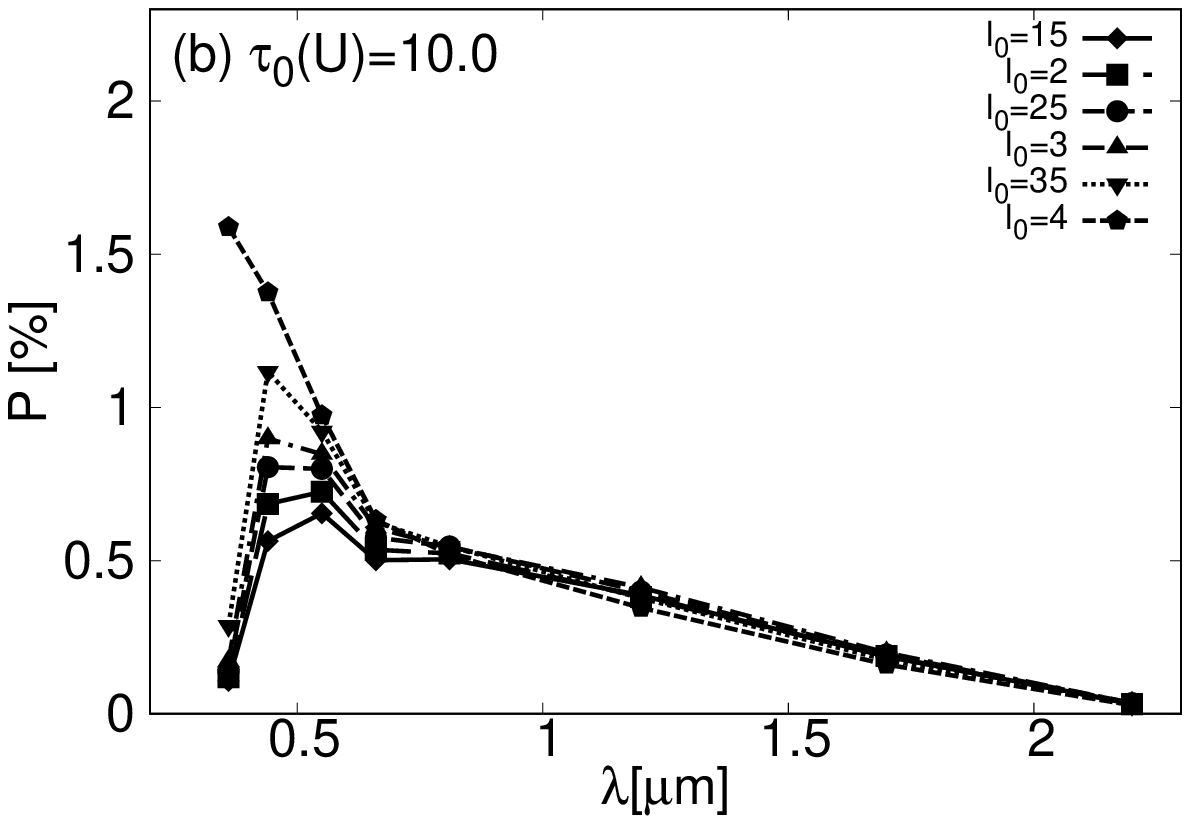}
  \caption{Same as Figure 10, but for various values of $l_{0}$ with $\theta_{\rm{obs}}=70^\circ$ (the LMC dust model).}
\end{figure*}

Figure 12 shows the wavelength dependence of the polarization for various dust models. In both case for $\tau_0(U)=1.0$ and $\tau_0(U)=10.0$, the overall behaviors are the same with the case in Figure 6: The wavelength dependence becomes stronger in the order of the C, MW, LMC, SMC dust models (see Section 3.1). However, the polarization is maximized at the $B$ or $V$ bands even in the case of $\tau_0(U)=10.0$ due to the wavelength dependence of the LCs.

\begin{figure*}[h]
  \includegraphics[scale=0.7]{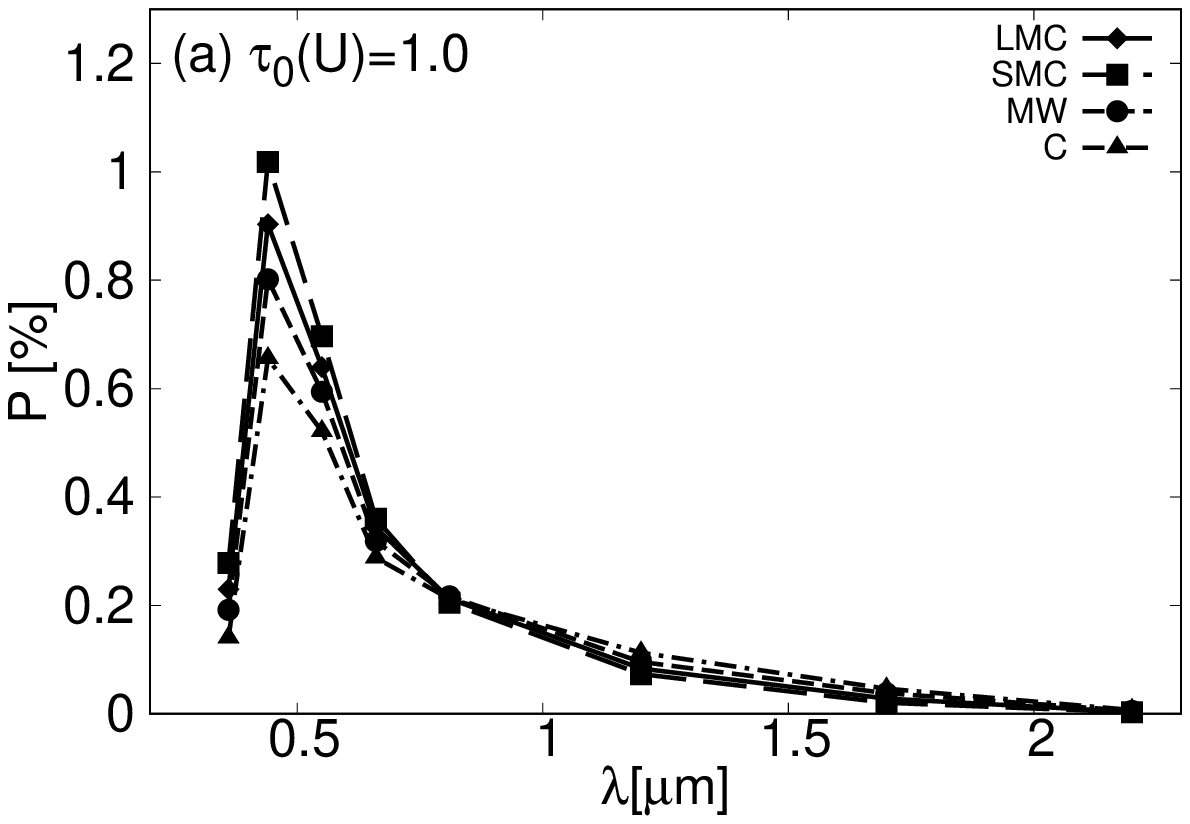}
  \includegraphics[scale=0.7]{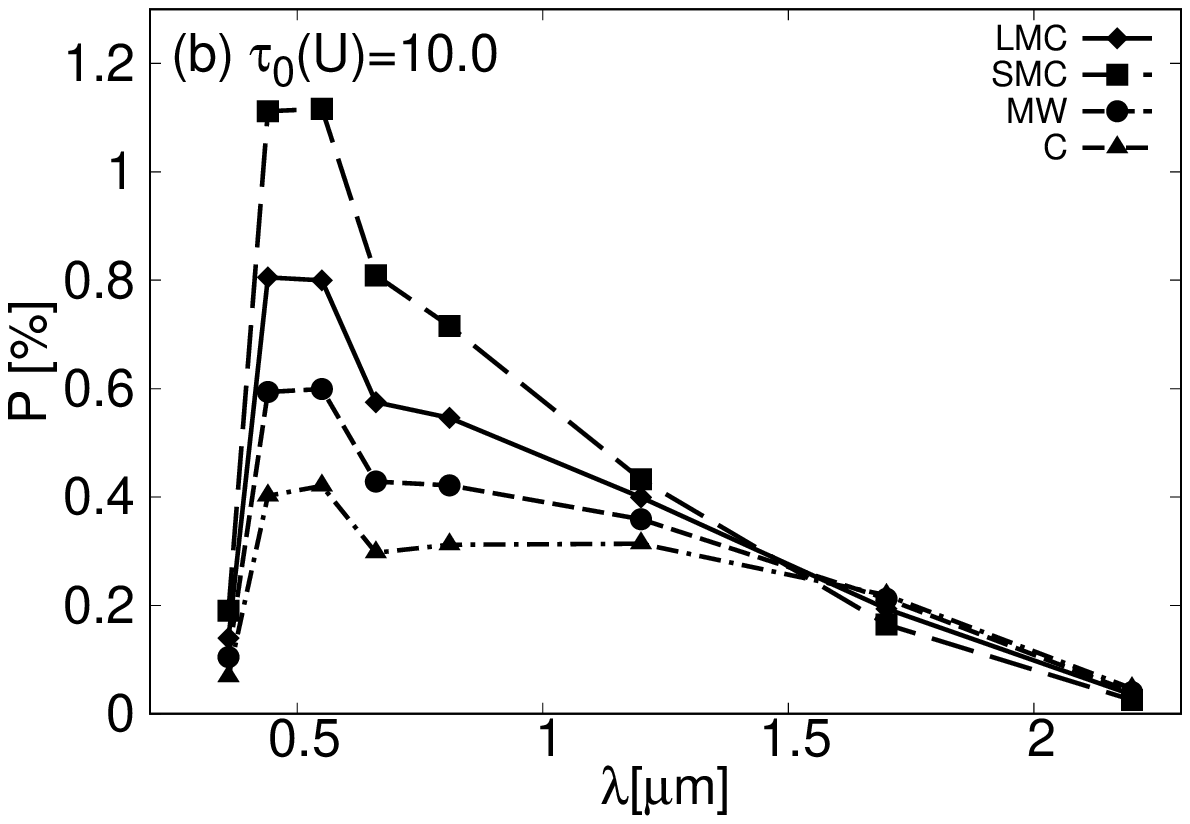}
  \caption{Same as Figure 6, but for the realistic LC model ($l_{0}=2.5 \times 10^{17}$ cm and $\theta_{\rm{obs}}=70^\circ$).}
\end{figure*}

\subsection{Systematic behaviors}
Here, we summarize the behavior of the polarization for different values of the tunable parameters. As an input SN light, we adopt the realistic LC model. There are then four tunable parameters in our model: the dust model, the optical depth of the blob ($\tau_0(U)$), the distance to the blob ($l_0$), and the viewing direction ($\theta_{\rm{obs}}$). As is shown above, the wavelength dependence of the polarization on the dust model is relatively simple; The dust model with more silicate grains instead of graphite grains makes the dependence stronger. For clearly understand the dependence on the other parameters ($\tau_0(U)$, $l_0$, $\theta_{\rm{obs}}$), we experimentally fit the calculated wavelength dependence by the following phenomenological function, which is inspired by the Serkowski law \citep[][]{Serkowski1975}:
\begin{eqnarray}
P(\lambda) = P_{\rm{max}}(B) f(\lambda),\\
f(\lambda) = \exp \left(-K \ln^2  \left(\frac{\lambda (B)}{\lambda} \right) \right),
\end{eqnarray}
where $P_{\rm{max}}(B)$ is a maximum value of the $B$-band polarization and $K$ is a parameter to express the steepness of the polarization curve. The shapes of the function $f(\lambda)$ for various $K$ are shown in Figure 13. 

\begin{figure}[h]
  \includegraphics[scale=0.7]{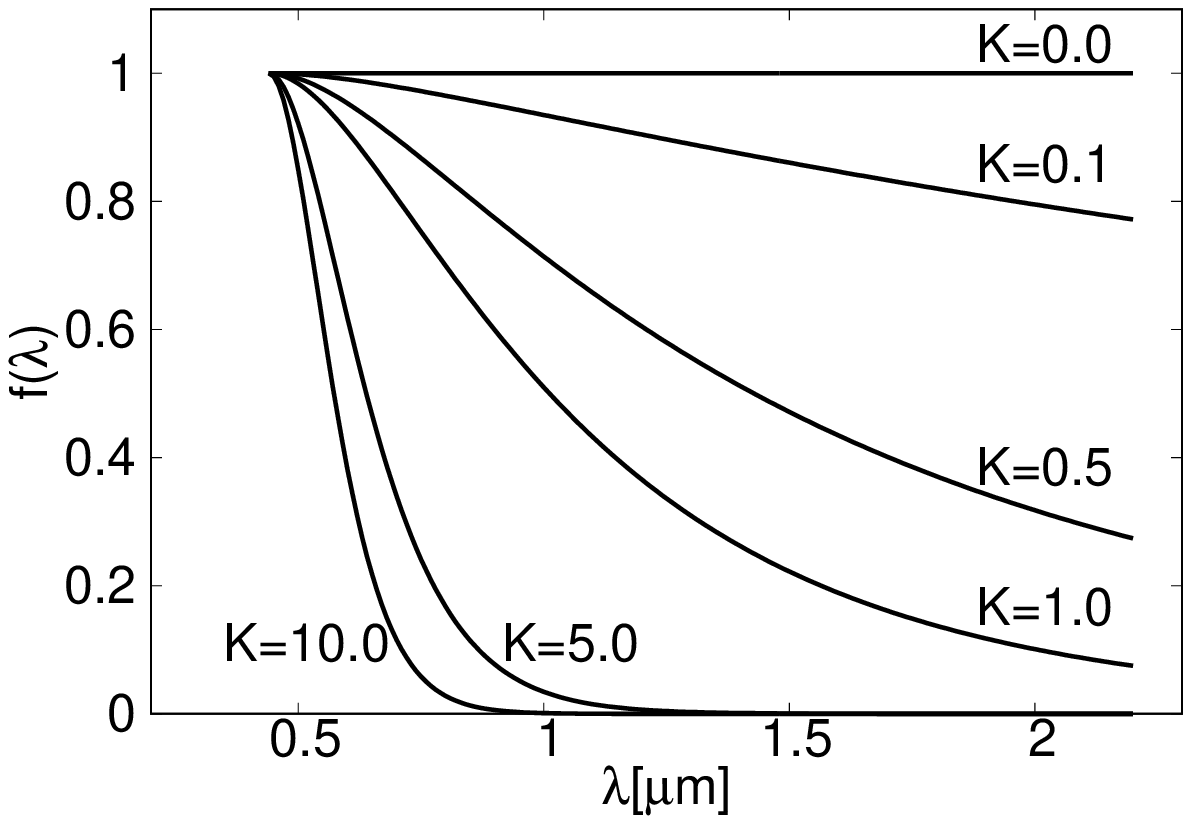}
  \caption{Function $f(\lambda)$.}
\end{figure}

The maximum value of the $B$-band polarization ($P_{\rm{max}}(B)$) and the steepness of the polarization curve ($K$) with the LMC dust model for various optical depth of the blob ($\tau_0(U)$), various distances to the blob ($l_0$) and various viewing directions ($\theta_{\rm{obs}}$) are shown in Figure 14. These are derived by comparing the calculated wavelength dependence with the function $f(\lambda)$, using only for the $B$, $V$, $R$ and $I$ bands. We do not use the calculated values for the $J$, $H$ and $K_s$ bands, because the values are the upper limits as is mentioned in Section 2. Figure 15 is the same as Figure 14, but showing the positions of each models with various optical depth of the blob ($\tau_0(U)$), various distances to the blob ($l_0$) and various viewing directions ($\theta_{\rm{obs}}$) in the $K$-$P_{\rm{max}}(B)$ plane. The maximum value ($P_{\rm{max}}(B)$) and the steepness ($K$) can be derived through multi-band polarimetric observations of SNe IIP. Thus, we can derive the best-fit parameters in the dust scattering model from the observations, and evaluate whether the parameters are consistent with those expected for an SN IIP.

\begin{figure*}[h]
  \includegraphics[scale=0.7]{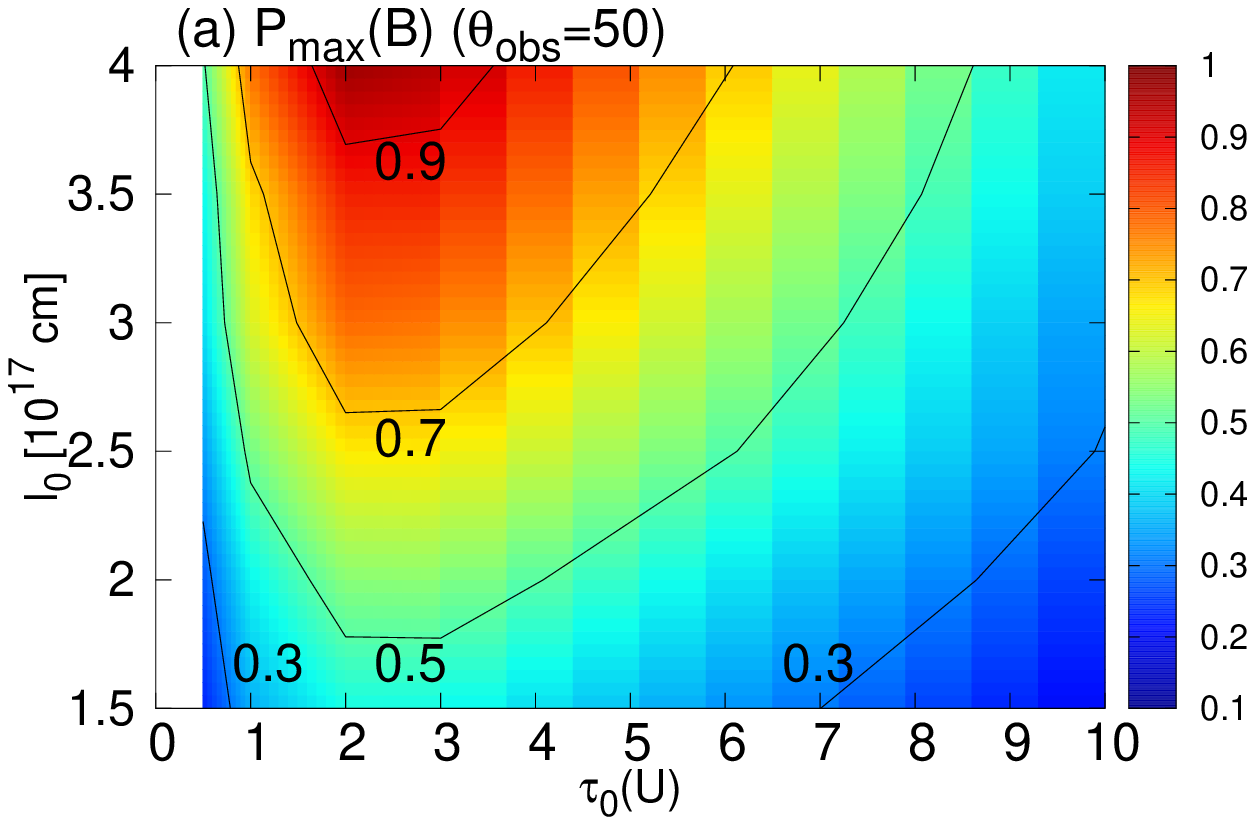}
  \includegraphics[scale=0.7]{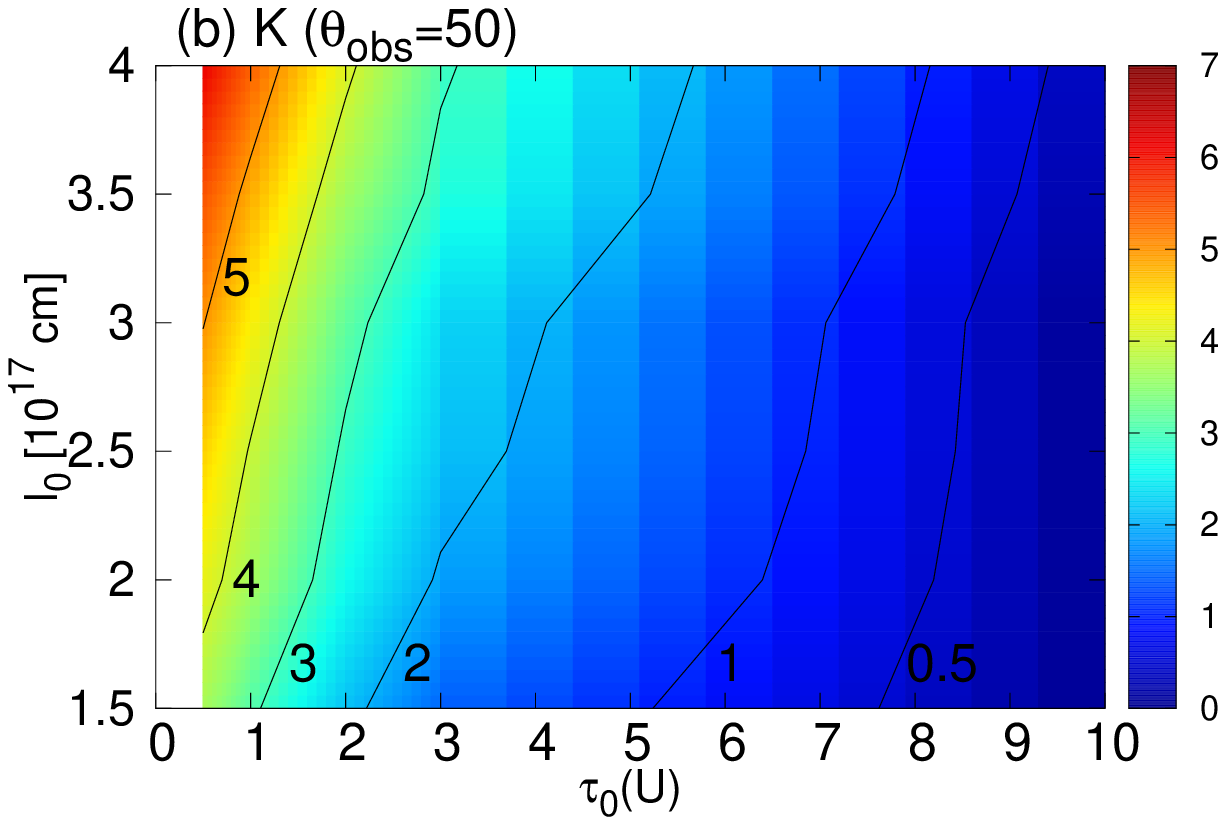}
  \includegraphics[scale=0.7]{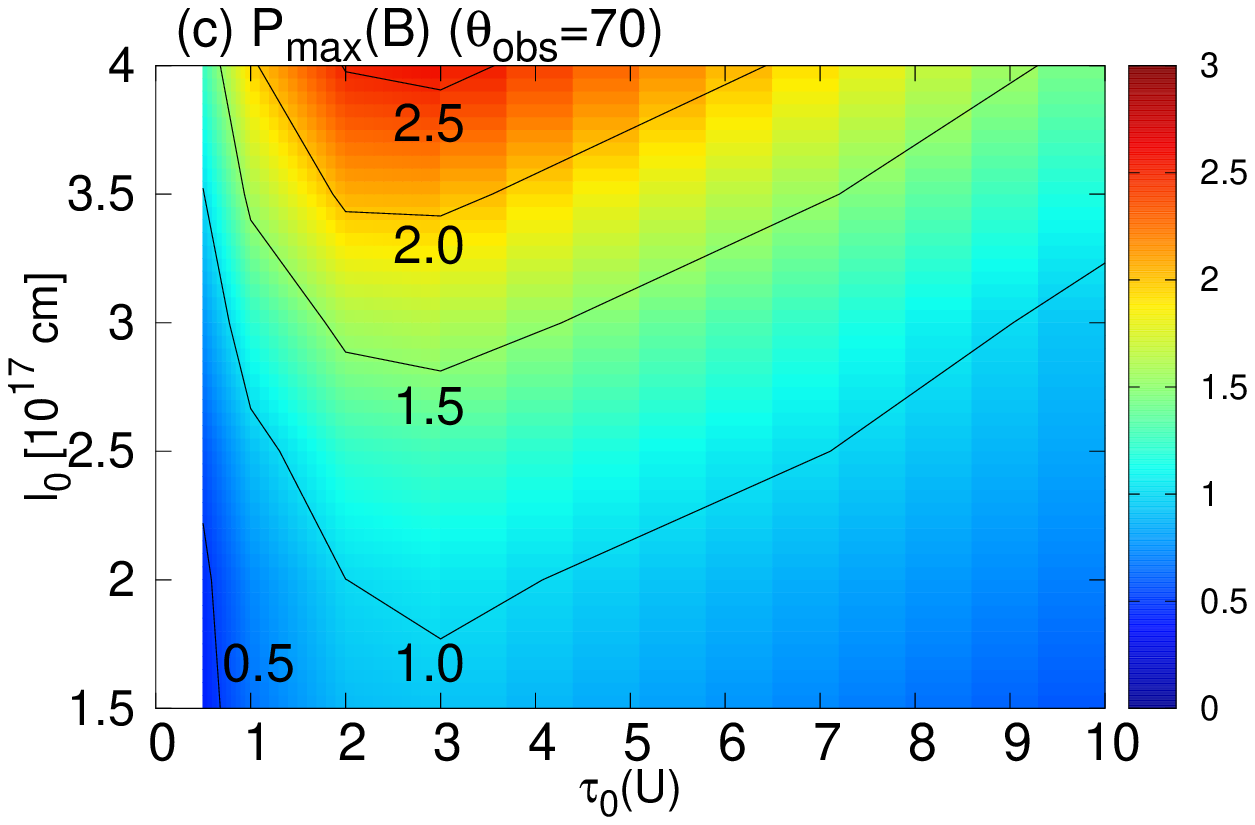}
  \includegraphics[scale=0.7]{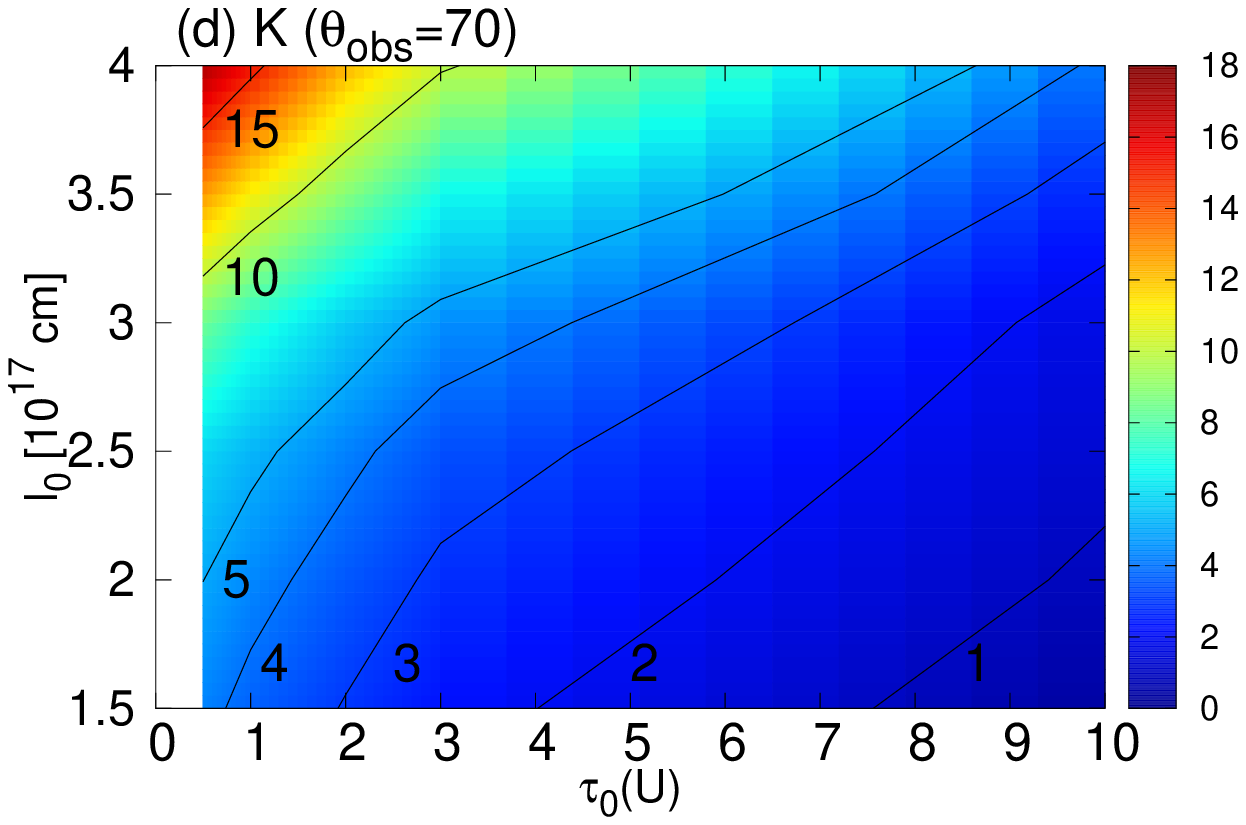}
  \includegraphics[scale=0.7]{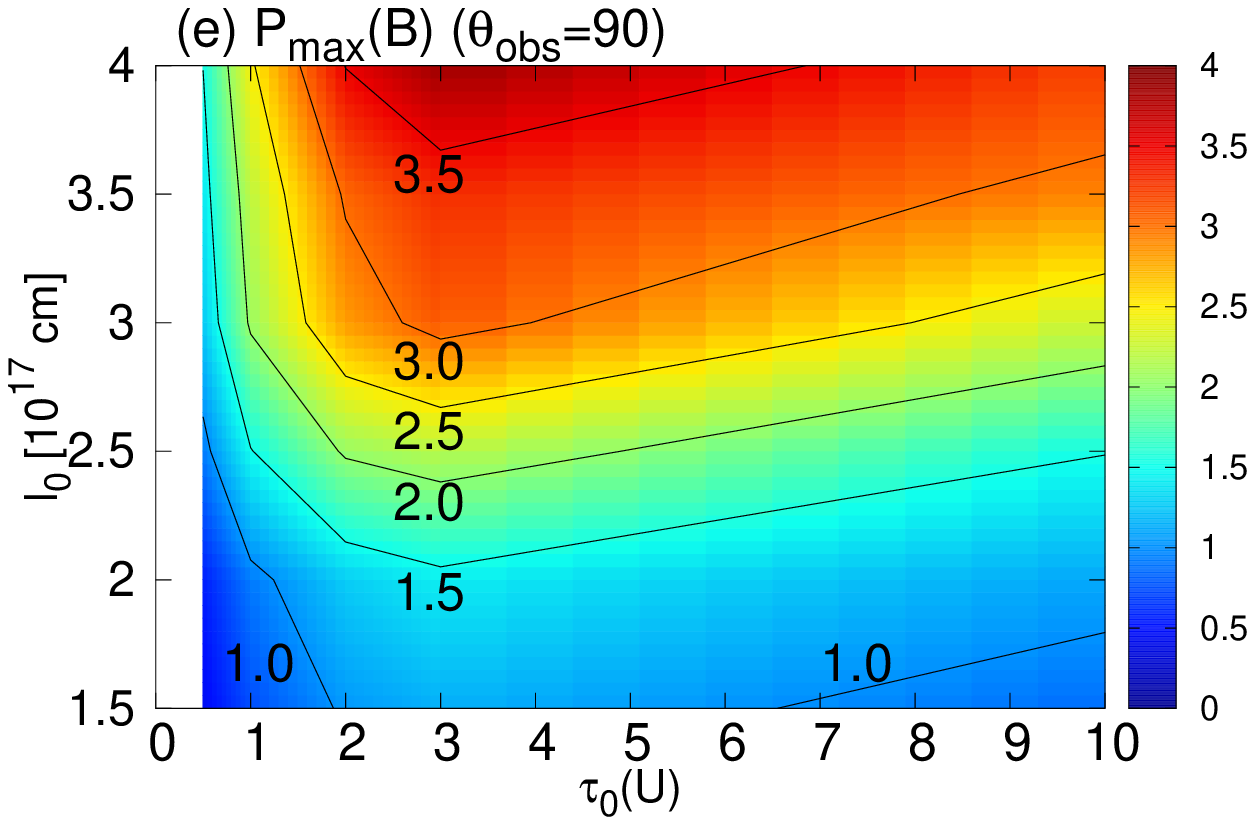}
  \includegraphics[scale=0.7]{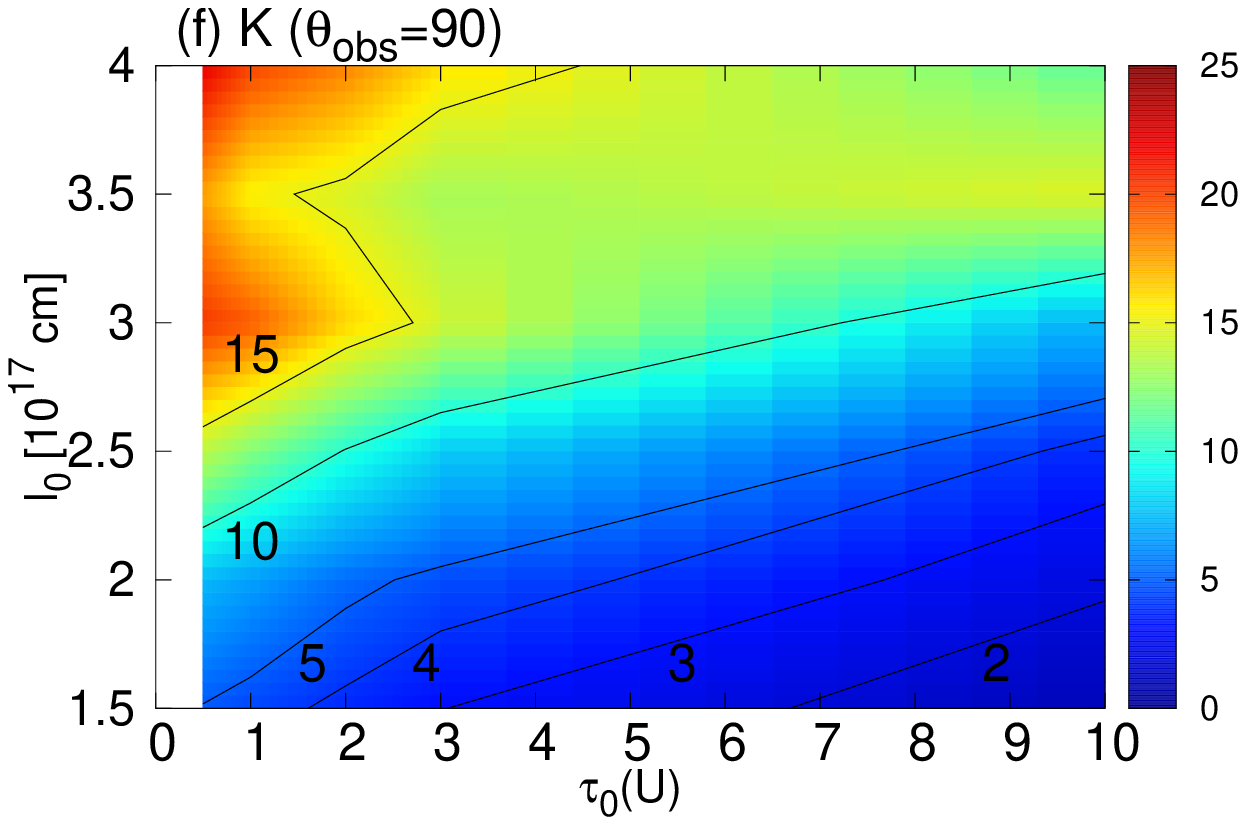}
  \caption{Values of $P_{\rm{max}}(B)$ and $K$ in the LMC dust model for various values of $\tau_0(U)$, $l_0$ and $\theta_{\rm{obs}}$.}
\end{figure*}

\begin{figure*}[h]
  \includegraphics[scale=0.7]{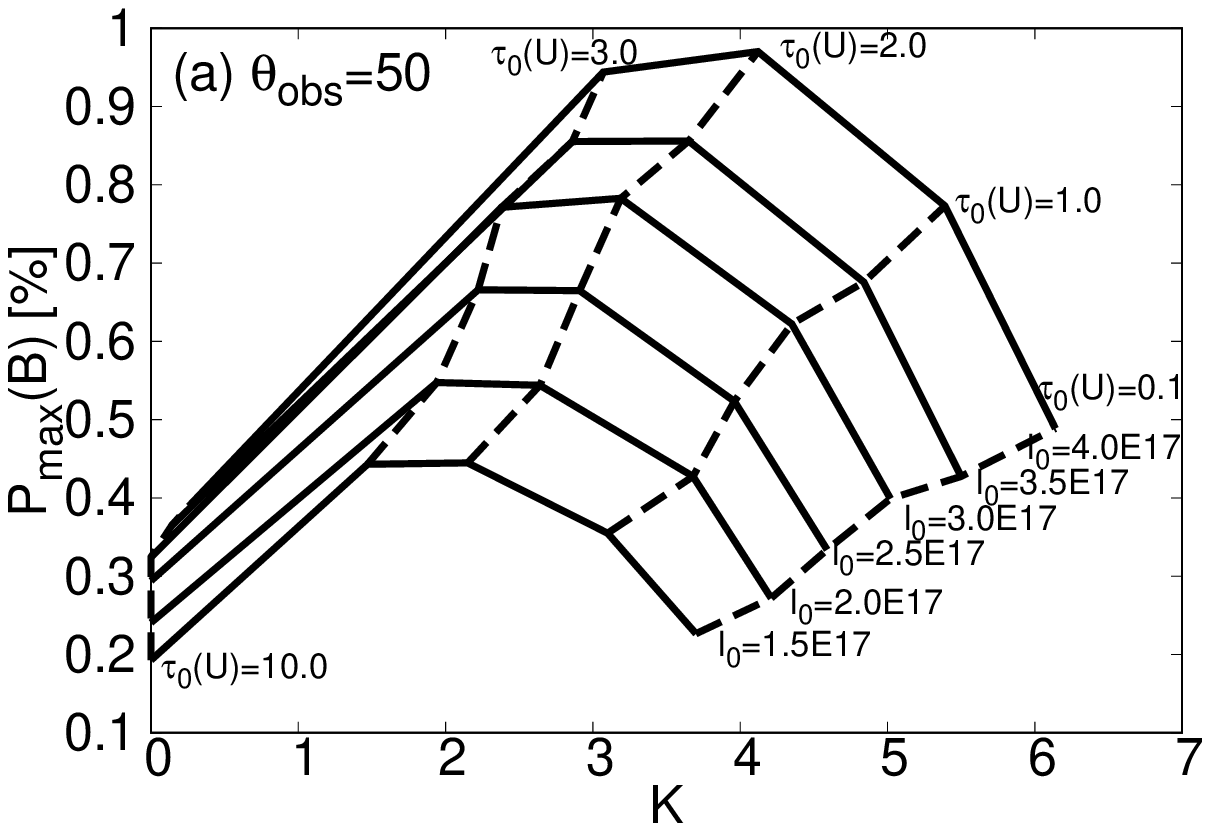}
  \includegraphics[scale=0.7]{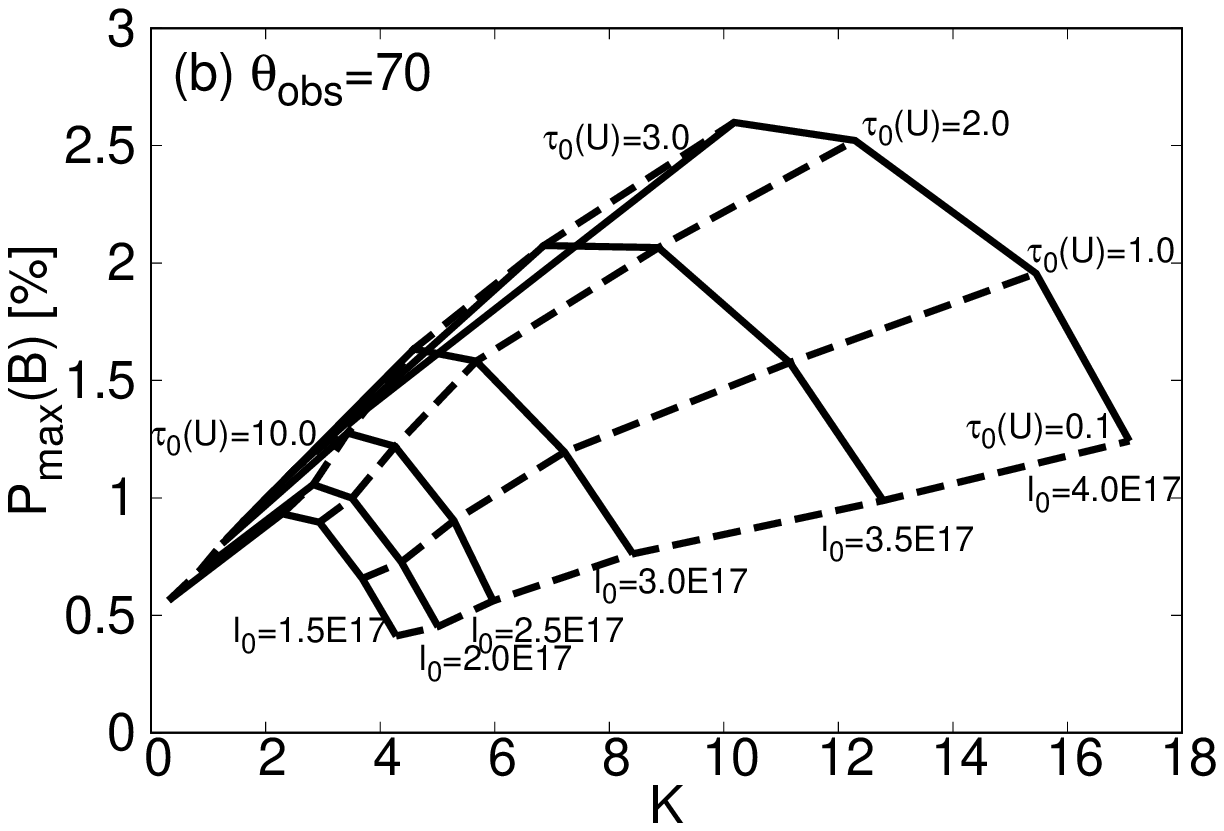}
  \includegraphics[scale=0.7]{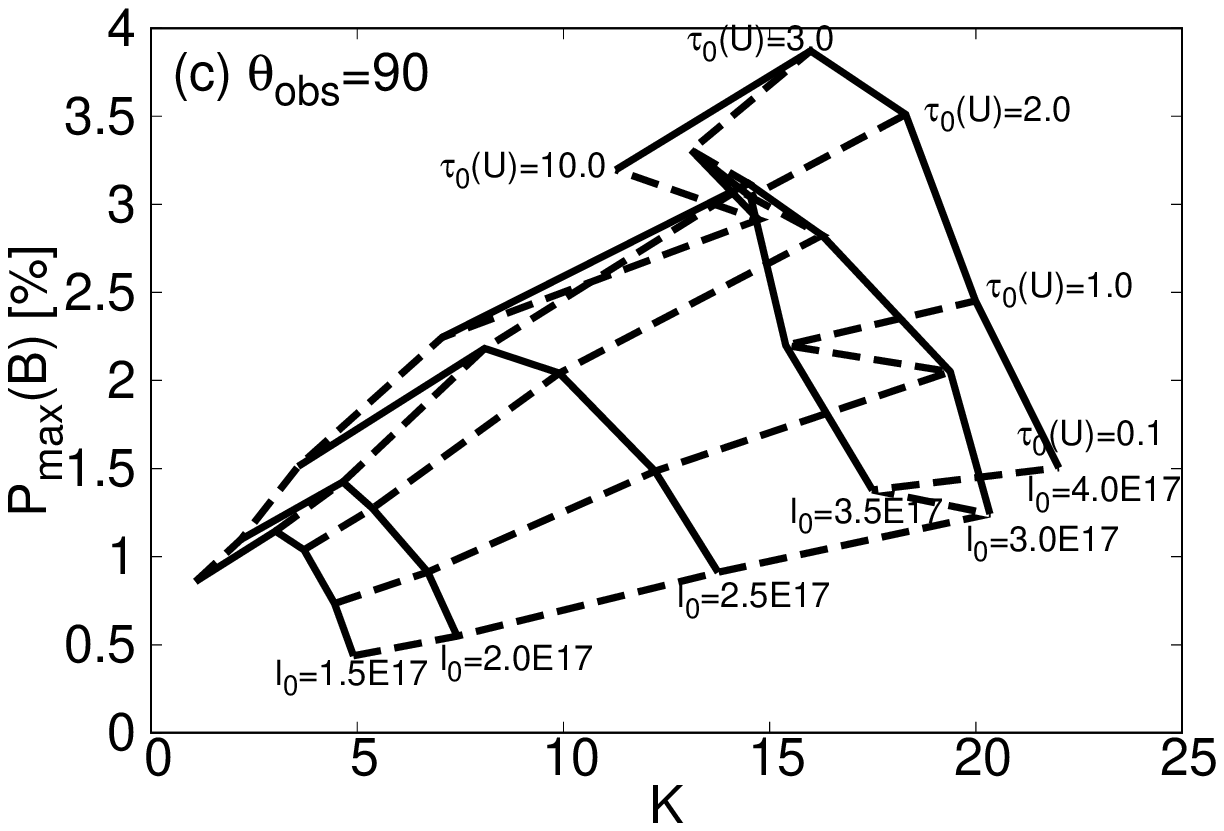}
  \caption{Same as Figure 14, but showing the positions of each model with various values of $\tau_0(U)$ and $l_0$ in the $K$-$P_{\rm{max}}(B)$ plane, toward various viewing directions.}
\end{figure*}

\section{Conclusions}
We have examined the wavelength dependence of the polarization in the dust scattering model. The predictions from our calculations in the dust scattering model are summarized as follows:
\begin{enumerate}
\item The $U$-band polarization emerges at an earlier phase with higher polarization degree than those in the other bands, though SNe IIP with the slower declines in the $U$ band like the Nugent's template may not show high polarization degrees. This is a unique feature that is distinguishable from the polarization due to an asymmetric ejecta (see Section 1). 
\item The information on the time and polarization degree when the $U$-band polarization is maximized is helpful to constrain the parameters in our model; This time corresponds to the light travel time, which is related to the distance to the blob ($l_0$) and the viewing direction ($\theta_{\rm{obs}}$). This polarization degree is connected mainly with the optical depth of the blob ($\tau_0(U)$) (and weakly with the dust model).
\item The polarization in the other optical bands (the $B$ to $I$ bands) is increased just after entering the nebular phase. At this phase, the polarization degree for the shorter wavelength is generally higher. This is another prediction that is different from those in the aspherical core model (see Section 1). 
\item This wavelength dependence is affected mainly by the following parameters in our model: The optical depth of the blob ($\tau_0(U)$), the distance to the blob ($l_0$) and the viewing direction ($\theta_{\rm{obs}}$) (and weekly by the dust model). Thus, these parameters can be derived from multi-band polarimetric observations of SNe IIP (Figure 15).
\item In the case of high optical depth of the blob ($\tau_0(U)$), the time evolution of the wavelength dependence is observed during the high polarization phase (from the beginning of the nebular phase through the delay time after the time).
\item The polarization degree in the near-infrared bands is quite low ($P < 0.5$ \%).
\end{enumerate}
With future multi-band polarimetric observations of SNe IIP, we can discuss the origin of the polarization of SNe IIP by comparing the observational data with the above features. We thus encourage multi-band polarimetric observations for SNe IIP.

\acknowledgments

Simulations were in part carried out on the PC cluster at Center for Computational Astrophysics, National Astronomical Observatory of Japan. The work has been supported by Japan Society for the Promotion of Science (JSPS) KAKENHI Grant 17J06373 (T.N.),17H02864 (K.M.) and15H02075, 16H02183 and 17H06363 (M.T.).


\end{document}